\documentclass[english,10pt,aps,prb]{revtex4}

\usepackage[T1]{fontenc}
\usepackage[latin9]{inputenc}
\usepackage{float}
\usepackage{amsmath}
\usepackage{graphicx}
\usepackage{setspace}
\usepackage{amssymb}
\usepackage{url}
\usepackage[breaklinks=false]{hyperref} 
\usepackage{indentfirst} 
\usepackage{subfigure}
\usepackage{footmisc} 



\usepackage{babel}

\begin{document}

\title{Quasiparticle scattering from vortices in d-wave superconductors I: Superflow contribution}

\author{Manas Kulkarni$^{1,2}$, Sriram Ganeshan$^{1}$, Adam C. Durst$^{1}$}

\affiliation{$^{1}$Department of Physics and Astronomy, Stony Brook University,
Stony Brook, NY 11794-3800}

\affiliation{$^{2}$Department of Condensed Matter Physics and Material Science,
Brookhaven National Laboratory, Upton, NY-11973}

\maketitle
In the vortex state of a d-wave superconductor, massless Dirac quasiparticles are scattered from magnetic vortices via a combination
of two basic mechanisms: effective potential scattering due to the superflow swirling about the vortices and Aharonov-Bohm scattering due
to the Berry phase acquired by a quasiparticle upon circling a vortex. In this paper, we study the superflow contribution by calculating the differential cross section for a quasiparticle scattering from the effective non-central potential of a single vortex. We solve the massless Dirac equation in polar coordinates and obtain the cross section via a partial wave analysis. We also present a more transparent Born-limit calculation and in this approximation we provide an analytic expression for the differential cross section. The Berry phase contribution to the quasiparticle scattering is considered in a separate paper.
\maketitle


\section{Introduction}
The low energy excitations of a $d$-wave superconductor are the Bogoliubov quasiparticles that reside in the vicinity of the four gap nodes at which the order parameter vanishes. The physics of these quasiparticles can be probed experimentally via a variety of low temperature electrical and thermal transport measurements. Since quasiparticles are part electron and part hole, their energy is well defined but their charge is not. Thus, it is thermal current that follows quasiparticle current. Since $T \ll T_{c}$, we are well within the superconducting state. Since $T \ll \Delta_{0}$ (the gap maximum), transport is dominated by quasiparticles excited in the vicinity of the gap nodes. (Quasiparticle dispersion is therefore given by the anisotropic Dirac spectrum, $E=(v_{f}^{2}k_{1}^{2}+v_{2}^{2}k_{2}^{2})^{1/2}$,where $v_{f}$ is the Fermi velocity, $v_{2}$ is the slope of the gap, and $k_{1}$ and $k_{2}$ are defined locally about each node. We shall choose our axes such that gap nodes are located at $\pm p_{F}\hat{\bf x}$ and $\pm p_{F}\hat{\bf y}$ in momentum space.) Furthermore, the temperatures of interest are low enough that sources of inelastic scattering are frozen out. This is what we mean by low temperature quasiparticle transport. In the mixed state, the remaining energy scales are the impurity scattering rate, $\Gamma_{0}$, the vortex scattering rate, $\Gamma_{v}$, and the temperature. We will focus on the weak-field high-temperature regime where $\Gamma_{v} \ll T$ and T is the dominant energy scale. In this regime, the quasiparticles responsible for transport are thermally generated rather than impurity-induced\cite{pa_lee_1993,graf_1996,durst_2000} or magnetic field-induced\cite{volovik_1993,kopnin,volovik_1997}.(Note that thermal transport in the opposite, low $T$, regime has been discussed frequently in the literature\cite{franz,vekhter,ashwin,vafek,Ye}.) In this high $T$ regime, the physical situation is relatively simple. Thermally excited quasiparticles carry the heat current. To understand the thermal transport, we need only understand how they scatter from magnetic vortices. Furthermore, since $H \ll H_{c2}$, the vortices are dilute, separated by distances large compared to the quasiparticle de Broglie wavelength. We can therefore learn a lot by considering the scattering of quasiparticles from a vortex and assuming that all such scattering events are independent.  We take the following path. In Sec.~\ref{sec:vortscat}, we develop our picture of a nodal quasiparticle scattering from a vortex. Then in Sec.~\ref{sec:BdG}, we consider the Bogoliubov-de Gennes equation, make a singular gauge transformation, and shift the origin of momentum space to the location of one of the gap nodes. The resulting problem is one of an (anisotropic) Dirac fermion scattering from an effective non-central potential (due to the superflow) in the presence of antiperiodic boundary conditions (a consequence of our gauge choice). After obtaining the quasiparticle current functional in Sec.~\ref{sec:qcurrent}, we define our vortex model in Sec.~\ref{sec:modelapprox} and make several important approximations. As far as scattering is concerned, the vortex has two parts: a circulating superflow and a Berry phase factor of (-1) acquired upon circling a vortex. In this paper, we consider scattering due to superflow current in a single vortex, without the Berry phase effect. (We consider the Berry phase contribution in a separate paper\cite{gkd}.) Furthermore, we neglect the anisotropy of the Dirac dispersion and take $v_{f}=v_{2}$. Although these approximations will prevent our results from being quantitatively accurate, we argue that our qualitative results reflect the essential physics of the problem. This notion is supported by Ref. \onlinecite{adamprl}, wherein this model, with these approximations, was used to derive a thermal Hall conductivity in good agreement with the qualitative features of the experimental results for $YBa_{2}Cu_{3}O_{6.99}$ by Zhang et al \cite{ong}. In Secs.~\ref{sec:angmom} and \ref{sec:radeq}, we expand the quasiparticle wave function in angular momentum eigenstates and obtain the resulting radial equations. We solve these equations outside the vortex in Sec.~\ref{sec:outside} and inside the vortex in Sec.~\ref{sec:inside}. In Sec.~\ref{sec:4node}, we match solutions and sum over all four gap nodes in order to obtain the single vortex scattering cross section. Results of our numerical calculations are presented in Sec.~\ref{sec:vortexresults}. In the same section (Sec.~\ref{sec:vortexresults}) we write a closed-form expression for the differential cross section using the more transparent Born approximation details of which are given in Appendix \ref{bornsc}. Conclusions are discussed in Sec \ref{conc}. Note that the results of this calculation were used in Ref.~\onlinecite{adamprl} (without the detailed derivation presented here) to calculate thermal transport coefficients.

\section{Single Vortex scattering of Bogoliubov quasiparticles}
\label{sec:vortscat}
\begin{figure}
\noindent \begin{centering}
\centerline{\resizebox{3in}{!}{\includegraphics{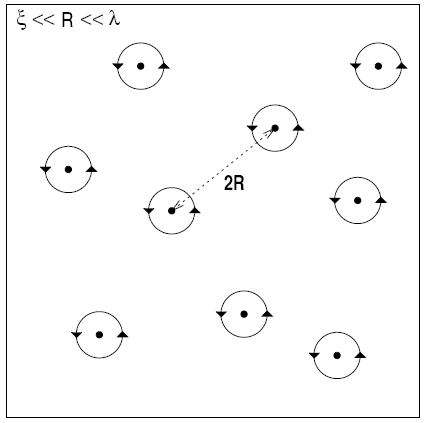}}}
\caption[Schematic depiction of vortices in the mixed state of
an extreme type~II superconductor.]
{Schematic depiction of vortices in the mixed state of
an extreme type~II superconductor. Vortices are pinned to randomly
distributed pinning centers and are separated by an average
inter-vortex distance, $2R$. Since $\xi \ll R \ll \lambda$,
vortex cores are point-like while the magnetic field profile
of each vortex is quite extended. Therefore, the magnetic field is
approximately uniform across the sample even though the vortices
are well separated.}
\label{fig:type2}
\par\end{centering}
\end{figure}

Consider the mixed state of a $d$-wave superconductor. Our picture is that of Fig.~\ref{fig:type2}. In the presence of a magnetic field ($H > H_{c1}$), magnetic vortices penetrate the sample (a 2D CuO$_{2}$ layer). Vortices are distributed randomly, pinned to local defects. The cuprates are extreme type II superconductors in which the coherence length, $\xi$, is much smaller than the penetration depth, $\lambda$. As a result, while the vortex cores may be well separated, the magnetic field profiles overlap significantly such that there is little variation in the magnetic field across the sample. We therefore adopt the extreme type II limit of $\xi \rightarrow 0$ and $\lambda \rightarrow \infty$ and take the magnetic field to be constant, ${\bf H} = H \hat{\bf z}$. In this limit, there are only two remaining length scales. The first, $1/k$, is set by the temperature such that $k \equiv E / \hbar v_{f} = k_{B} T / \hbar v_{f}$. For anisotropic Dirac nodes, $k = \sqrt{k_{1}^{2} + k_{2}^{2} (v_{2}/v_{f})^{2}}$. Note that in the isotropic limit, this reduces to the magnitude of the quasiparticle momentum (measured from a node). The second length scale,
$R$, is half of the average distance between vortices. With one flux quantum per vortex, $H \pi R^{2} = \Phi_{0} = hc/2e$, so we define $R \equiv \sqrt{\hbar c / e H}$. In terms of $R$, we can define the (2D) density of vortices to be $n_{v} = H/\Phi_{0} = 1/\pi R^{2}$. Note that the ratio of these two length scales yields
\begin{equation}
kR = \frac{k_{B} T}{\hbar v_{f}} \sqrt{\frac{\hbar c}{e H}}
= \frac{\gamma T}{\sqrt{H}}
\label{eq:kR}
\end{equation}
which is the inverse of the argument of the Simon-Lee scaling functions\cite{simon}.

\begin{figure}
   \includegraphics[width=6cm]{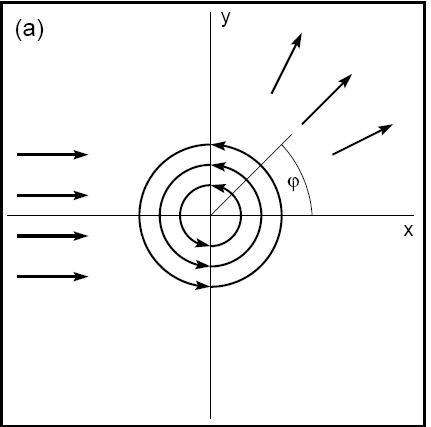}\qquad\qquad\includegraphics[width=6cm]{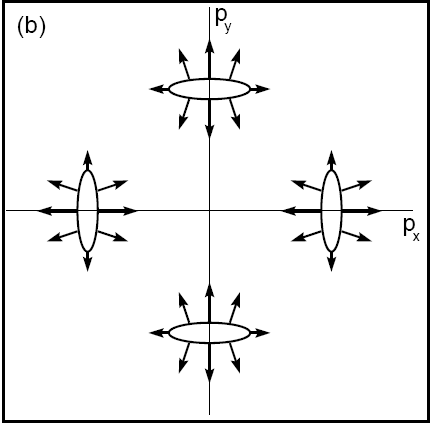}
  \caption[Schematic depictions of momentum space and coordinate space
for a single vortex in a $d$-wave superconductor.]
{Schematic depictions of momentum space and coordinate space.
(a) Quasiparticle scattering from a single vortex. An incident
plane wave scatters from an origin-centered vortex into an
outgoing radial wave.(b) Gap nodes in the Brillouin zone of a CuO$_{2}$ layer.
Ellipses denote surfaces of equal energy. Arrows represent the
quasiparticle group velocity in the vicinity of each node.}
\label{fig:BZrotated}
\end{figure}

We wish to consider a quasiparticle, in a state with a particular energy and current, scattering from a magnetic vortex into another state, of the same energy, with a different current. At low temperatures, quasiparticles are excited in the vicinity of the four gap nodes in momentum space. The nodal structure of the Brillouin zone is depicted in Fig.\ref{fig:BZrotated}(b). Following Simon and Lee \cite{simon}, we have defined coordinate axes, rotated by $45^{\circ}$ from the usual crystal axes, such that the gap nodes are located on-axis at $\pm p_{F} \hat{\bf x}$ and
$\pm p_{F} \hat{\bf y}$. In the neighborhood of a node, the quasiparticle excitation energy is given by the anisotropic Dirac spectrum, $E_{k}=\sqrt{v_{f}^{2}k_{1}^{2}+v_{2}^{2}k_{2}^{2}}$, where the $k_{1}$ and $k_{2}$ axes are defined, respectively, to be perpendicular and parallel to the local Fermi surface. The surfaces of equal energy are therefore ellipses centered about each of the nodes. The quasiparticle current is given by the group velocity ${\bf v}_{G} = {\bf \nabla}_{k} E_{k}$, directed outward from the node centers, perpendicular to the ellipses of constant energy.

In coordinate space, the quasiparticle is represented as an incident plane wave. The single ($hc/2e$) vortex is centered at the origin and surrounded by a circulating supercurrent. Due to the interaction between the quasiparticle
and the vortex, the quasiparticle is scattered into an outgoing radial wave. The situation is depicted schematically in Fig.~\ref{fig:BZrotated}(a). By solving the Bogoliubov-de Gennes equation for the quasiparticle wave functions
and considering the nature of the quasiparticle current before and after such a scattering event, we will compute the quasiparticle cross section for scattering from a vortex in a $d$-wave superconductor. Our analysis shall be similar in spirit to that conducted for the $s$-wave case by Cleary\cite{cleary-1968,cleary-1970}, who extended the work of Caroli, de Gennes, and Matricon \cite{matricon} (on vortex core bound states) to the problem of quasiparticle scattering from a vortex in an $s$-wave superconductor.

\section{Bogoliubov-de Gennes Equation}
\label{sec:BdG}
Consider the Bogoliubov-de Gennes (BdG) equation for a $d$-wave superconductor in the presence of a constant perpendicular magnetic field, ${\bf A}=\frac{1}{2}Hr\hat{\bf \phi}$, and with an order parameter that winds once about the origin, $\Delta({\bf r})=\Delta_{0}e^{i\phi}$:
\begin{equation}
H^{\prime} \Psi = E \Psi \;\;\;\;\;\;\;\;
H^{\prime} =
\left( \begin{array}{cc} \hat{H}_{e}^{\prime} & \hat{\Delta}^{\prime} \\
\hat{\Delta}^{\prime *} & -\hat{H}_{e}^{\prime *} \end{array} \right)
\label{eq:BdG}
\end{equation}
\begin{equation}
\hat{H}_{e}^{\prime} = \frac{1}{2m} \left( {\bf p}-\frac{e}{c}{\bf A}
\right)^{2} - E_{F}
\label{eq:He}
\end{equation}
\begin{equation}
\hat{\Delta}^{\prime} = \frac{1}{p_{F}^{2}}
\{\hat{p_{x}},\{\hat{p_{y}},\Delta({\bf r})\}\}
-\frac{i}{4p_{F}^{2}} \Delta({\bf r})
(\partial_{x} \partial_{y} \phi)
\label{eq:DeltaOp}
\end{equation}
Here ${\bf p}=-i\hbar {\bf \nabla}$, $\{a,b\}=(ab+ba)/2$, $E$ is the quasiparticle energy, and the form of the gap operator is that required to provide $d$-wave symmetry \cite{vafek,simon}. Entangled within the complex differential form of the gap operator, $\hat{\Delta}$, is the fact that upon circling an $hc/2e$ vortex, the quasiparticle acquires a Berry phase factor of (-1). The Hamiltonian would be simplified if we could effectively strip the gap function, $\Delta({\bf r})$, of its phase. This is accomplished via the application of the singular gauge transformation
\begin{equation}
U = \left( \begin{array}{cc} e^{-i\phi/2} & 0 \\
0 & e^{i\phi/2} \end{array} \right)
\;\;\;\;\;\;\;\; \Phi({\bf r}) = U^{-1} \Psi({\bf r})
\;\;\;\;\;\;\;\; H = U^{-1} H^{\prime} U .
\label{eq:gaugetrans}
\end{equation}
In this gauge, known as the Anderson gauge,
\begin{equation}
H \Phi = E \Phi
\label{eq:transBdG}
\end{equation}
\begin{equation}
H = \tau_{3} \frac{v_{f}}{2p_{F}}
\left[ ({\bf p} + \tau_{3} {\bf P}_{s})^{2} - p_{F}^{2} \right]
+ \tau_{1} \frac{v_{2}}{2p_{F}}
\left[ 2 p_{x} p_{y} \right]
\label{eq:transH}
\end{equation}
where
\begin{equation}
{\bf P}_{s}({\bf r}) =
\frac{\hbar}{2} {\bf \nabla} \phi - \frac{e}{c} {\bf A} =
\frac{\hbar}{2} \left( \frac{1}{r} - \frac{r}{R^{2}} \right) \hat{\bf \phi}
\label{eq:ps}
\end{equation}
is the gauge invariant superfluid momentum (superflow), $v_{f}=p_{F}/m$, $v_{2}=\Delta_{0}/p_{F}$, and $R \equiv \sqrt{\hbar c / e H}$. Although this form of the Bogoliubov-de Gennes equation is much simplified, the boundary conditions have become more complicated. While the original wave function was defined with periodic boundary conditions, $\Psi(r,\phi)=\Psi(r,\phi+2\pi)$, the transformed wave function is not single-valued and has antiperiodic boundary conditions, $\Phi(r,\phi)=-\Phi(r,\phi+2\pi)$. Hence, we have introduced a branch cut such that with each trip around
the origin, the wave function changes sign. In effect, the Berry phase contribution has been extracted from the
Hamiltonian and encoded in the antiperiodic boundary conditions imposed on the wave function. Note that when considering a system of many vortices, the resulting sea of branch cuts can be problematic. In response, Franz and Tesanovic
 introduced an alternate gauge transformation better suited to the vortex lattice \cite{vafek-ashot,franztesano}. However, for the single vortex case that we consider, the Anderson gauge described above is most convenient. All observables, such as differential cross section or transport coefficients, are independent of gauge choice. We would like to emphasize that the Berry phase effect is not an artifact of our gauge choice.  It is physical, and an intrinsic part of the original Bogoliubov de Gennes equations.  In the untransformed Hamiltonian, which has periodic boundary conditions, it resides in the complex differential form of the gap operator.  The gauge transformation drastically simplifies the Hamiltonian at the cost of introducing antiperiodic boundary conditions.  Thus, effectively, the Berry phase contribution is extracted from the Hamiltonian and moved to the boundary conditions.

Recall that the low energy excitations of a $d$-wave superconductor are concentrated about the four nodal points in momentum space where the gap vanishes. We can therefore further simplify our Hamiltonian by shifting the origin of momentum space to the location of one of the nodes. Shifting to node 1
\begin{equation}
p_{x} \rightarrow p_{F} + p_{x} \;\;\;\;\;\;\;\; p_{y} \rightarrow p_{y}
\label{eq:pshift}
\end{equation}
we find that
\begin{equation}
H = H_{D} + H_{C}
\label{eq:shiftedH}
\end{equation}
\begin{equation}
H_{D} = v_{f} \left[ p_{x} \tau_{3} + \alpha p_{y} \tau_{1} + P_{sx} \right]
\label{eq:HDirac}
\end{equation}
\begin{equation}
H_{C} = \frac{v_{f}}{2p_{F}} \left[ (p^{2} + P_{s}^{2}) \tau_{3}
+ 2 {\bf P}_{s} \cdot {\bf p} + \alpha 2 p_{x} p_{y} \tau_{1} \right]
\label{eq:Hcurve}
\end{equation}
where $\alpha = v_{2}/v_{f}$ and we have used the fact that ${\bf P}_{s} = P_{s}(r) \hat{\bf \phi}$ to commute ${\bf p}$ with ${\bf P}_{s}$. Here $H$ is written as the sum of a linear (Dirac) Hamiltonian, $H_{D}$, and a quadratic (curvature) Hamiltonian, $H_{C}$, to emphasize that the second term is smaller than the first by a factor of $E/E_{F}$. We will focus on the dominant term, $H_{D}$ , in the remainder of this paper. In what follows we shall seek solutions to the Bogoliubov-de Gennes equation developed above. However, we must first consider the nature of quasiparticle current in a $d$-wave superconductor.

\section{Quasiparticle Current}
\label{sec:qcurrent}
In order to calculate the cross section for a quasiparticle scattering from a vortex, we must be able to write down the currents corresponding to the incident and scattered wave functions. Since the incident and scattered currents will be considered in the far field where the quasiparticles are free, we wish to determine the quasiparticle current as a functional of $\Phi$ for $P_{s}=0$. Setting $P_{s}=0$ in Eq.~(\ref{eq:transH}) we find that the BdG Hamiltonian becomes
\begin{equation}
H = \left( \begin{array}{cc} \hat{H}_{e} & \hat{\Delta} \\
\hat{\Delta}^{*} & -\hat{H}_{e}^{*} \end{array} \right)
\;\;\;\;\;\;\;\; \hat{H}_{e} = - \frac{v_{f}}{2p_{F}} \nabla^{2} - E_{F}
\;\;\;\;\;\;\;\; \hat{\Delta} = - \frac{v_{2}}{2p_{F}}
2 \partial_{x} \partial_{y}
\label{eq:transHps0}
\end{equation}
Following Ref.~\onlinecite{deGennes}, we can write down a mean-field Hamiltonian in terms of the real-space electron creation/annihilations operators, $\Psi_{\alpha}^{\dagger}$ and $\Psi_{\alpha}$.
\begin{equation}
H_{eff} = \int d^{2}r \left[
\Psi_{\uparrow}^{\dagger} \hat{H}_{e} \Psi_{\uparrow} +
\Psi_{\downarrow}^{\dagger} \hat{H}_{e} \Psi_{\downarrow} +
\hat{\Delta} \Psi_{\uparrow}^{\dagger} \Psi_{\downarrow}^{\dagger} +
\hat{\Delta}^{*} \Psi_{\downarrow} \Psi_{\uparrow} \right]
\label{eq:Heff}
\end{equation}
Since quasiparticles carry well-defined spin (and heat) the quasiparticle current is equal to the spin current divided by the quasiparticle spin. The spin density operator is expressed as
\begin{equation}
\hat{\rho}^{s}({\bf r}) = s \left[
\Psi_{\uparrow}^{\dagger}({\bf r}) \Psi_{\uparrow}({\bf r}) -
\Psi_{\downarrow}^{\dagger}({\bf r}) \Psi_{\downarrow}({\bf r}) \right]
\label{eq:rhosop}
\end{equation}
where $s \equiv 1/2$ and the spin current operator is, in turn, determined via
\begin{equation}
{\bf \nabla} \cdot \hat{\bf j}^{s}({\bf r}) = - \dot{\hat{\rho^{s}}}
= i \left[ \hat{\rho}^{s}({\bf r}) , H_{eff} \right] .
\label{eq:Jsop}
\end{equation}
Making use of the fermionic anticommutation relations,
\begin{equation}
\left\{ \Psi_{\alpha}({\bf r}) , \Psi_{\beta}({\bf r^{\prime}}) \right\} =
\left\{ \Psi_{\alpha}^{\dagger}({\bf r}) ,
\Psi_{\beta}^{\dagger}({\bf r^{\prime}}) \right\} = 0
\;\;\;\;\;\;\;\;
\left\{ \Psi_{\alpha}^{\dagger}({\bf r}) ,
\Psi_{\beta}({\bf r^{\prime}}) \right\} = \delta_{\alpha\beta}
\delta({\bf r} - {\bf r^{\prime}})
\label{eq:anticom}
\end{equation}
we find that
\begin{equation}
\left[ \hat{\rho}^{s}({\bf r}) , H_{eff} \right] =
-\frac{s}{2p_{F}} \sum_{\alpha} \left[
v_{f} \eta_{\alpha} \Psi_{\alpha}^{\dagger} \nabla^{2} \Psi_{\alpha} +
v_{2} \Psi_{\alpha}^{\dagger} 2 \partial_{x} \partial_{y} \Psi_{\bar{\alpha}}
- \mbox{h.c.} \right]
\label{eq:comrhoH}
\end{equation}
where $\eta_{\alpha} \equiv \pm 1$. Noting that each of these terms can be manipulated into the form of a divergence and making use of Eq.~(\ref{eq:Jsop}) we can write the spin current operator as
\begin{equation}
\hat{\bf j}^{s} = \frac{s}{2ip_{F}} \sum_{\alpha} \left[
v_{f} \eta_{\alpha} \Psi_{\alpha}^{\dagger} {\bf \nabla} \Psi_{\alpha} +
v_{2} \Psi_{\alpha}^{\dagger} {\bf \nabla}_{\perp} \Psi_{\bar{\alpha}}
- \mbox{h.c.} \right]
\label{eq:Jsop1}
\end{equation}
where ${\bf \nabla} = \partial_{x} \hat{\bf x} + \partial_{y} \hat{\bf y}$
and ${\bf \nabla}_{\perp} \equiv \partial_{y} \hat{\bf x}
+ \partial_{x} \hat{\bf y}$.
The real-space creation/annihilation operators,
$\Psi_{\alpha}^{\dagger}$ and $\Psi_{\alpha}$, can be expressed
as a weighted sum of Bogoliubov operators,
$\gamma_{n\alpha}^{\dagger}$ and $\gamma_{n\alpha}$, via
\begin{equation}
\Psi_{\uparrow}({\bf r}) = \sum_{n} \left(
\gamma_{n\uparrow} u_{n}({\bf r}) -
\gamma_{n\downarrow}^{\dagger} v_{n}^{*}({\bf r}) \right)
\;\;\;\;\;\;\;\;
\Psi_{\downarrow}({\bf r}) = \sum_{n} \left(
\gamma_{n\downarrow} u_{n}({\bf r}) +
\gamma_{n\uparrow}^{\dagger} v_{n}^{*}({\bf r}) \right)
\label{eq:real2bogol}
\end{equation}
where $u_{n}$ and $v_{n}$ are particle and hole wave functions for state $n$. Plugging these forms into the spin current operator and evaluating for state $\ell$ and spin $\sigma$ yields an expression for the spin current as a functional of $u_{\ell}$ and $v_{\ell}$.
\begin{equation}
{\bf j}^{s} = \langle \ell \sigma | \hat{\bf j}^{s} | \ell \sigma \rangle
= \frac{s\eta_{\sigma}}{2ip_{F}} \left[
v_{f} (u_{\ell}^{*} {\bf \nabla} u_{\ell} -
v_{\ell}^{*} {\bf \nabla} v_{\ell}) +
v_{2} (u_{\ell}^{*} {\bf \nabla}_{\perp} v_{\ell} +
v_{\ell}^{*} {\bf \nabla}_{\perp} u_{\ell}) - \mbox{c.c.} \right]
\label{eq:Js}
\end{equation}
Dropping the eigenstate label, dividing by the spin $s\eta_{\sigma}$, and writing the result in terms of the particle-hole 2-vector
\begin{equation}
\Phi = \left( \begin{array}{c} u \\ v \end{array} \right) \;\;\;\;\;\;\;\;
\Phi^{\dagger} = \left( u^{*} , v^{*} \right)
\label{eq:Phidefuv}
\end{equation}
we obtain the quasiparticle current
\begin{equation}
{\bf j} = \frac{1}{p_{F}} \mbox{Im} \left[
v_{f} \Phi^{\dagger} \tau_{3} {\bf \nabla} \Phi +
v_{2} \Phi^{\dagger} \tau_{1} {\bf \nabla}_{\perp} \Phi \right] .
\label{eq:JPhi}
\end{equation}
As in the preceding section, it is convenient to shift the origin of momentum space to a nodal point. Shifting to node 1 yields
\begin{equation}
{\bf j} = {\bf j}_{D} + {\bf j}_{C}
\label{eq:shiftedJ}
\end{equation}
\begin{equation}
{\bf j}_{D} = v_{f} \Phi^{\dagger} (\tau_{3} \hat{\bf x}
+ \alpha \tau_{1} \hat{\bf y}) \Phi
\label{eq:JDirac}
\end{equation}
\begin{equation}
{\bf j}_{C} = \frac{v_{f}}{p_{F}} \mbox{Im} \left[
\Phi^{\dagger} (\tau_{3} \hat{\bf x}
+ \alpha \tau_{1} \hat{\bf y}) \frac{\partial \Phi}{\partial x} +
\Phi^{\dagger} (\tau_{3} \hat{\bf y}
+ \alpha \tau_{1} \hat{\bf x}) \frac{\partial \Phi}{\partial y} \right]
\label{eq:Jcurve}
\end{equation}
where $\alpha=v_{2}/v_{f}$.
Once again, the second (curvature) term is smaller than the first (Dirac) term by a factor of $E/E_{F}$ and shall be neglected in what follows.

\section{Model and Approximations for single vortex}
\label{sec:modelapprox}
Given the BdG Hamiltonian and the quasiparticle current functional, we are ready to consider a model of quasiparticle scattering from a single vortex. In reality, each quasiparticle encounters a sea of vortices separated by distances on the order of $2R=\sqrt{4\hbar c/eH}$. The superflow circulating around each vortex is peaked near the vortex centers but falls off slowly and overlaps in the regions between vortices. The total flux through the sample is equal to one ($hc/2e$) flux quantum per vortex but is distributed smoothly since the penetration depth is much larger than the inter-vortex distance. In order to model this situation via single vortex scattering, we approximate the effect of neighboring vortices by cutting off the superflow distribution about our single vortex at a distance $R$ from its center. By construction, the flux through this circle is exactly one flux quantum. Thus, the superfluid momentum takes the form of Eq.~(\ref{eq:ps}) inside the vortex ($r<R$) and is zero outside the vortex ($r>R$) \begin{equation}
{\bf P}_{s}({\bf r}) =
\frac{\hbar}{2} \left( \frac{1}{r} - \frac{r}{R^{2}} \right)
\theta(R-r) \hat{\bf \phi} .
\label{eq:psmodel}
\end{equation}
Note that at the vortex edge ($r=R$), the phase gradient and vector potential terms cancel and the superflow goes continuously to zero. We shall consider a quasiparticle plane wave incident upon such a vortex and calculate the scattering cross section. Before going forward, however, we make the following additional approximations.

First, recall from Sec.~\ref{sec:BdG} that, with our gauge choice, the wave function must obey antiperiodic boundary conditions such that it changes sign each time the quasiparticle winds around the vortex. This is the Berry phase contribution to the problem and it has the interference effect of an Aharonov-Bohm $\pi$-flux at the origin. Thus, even in the absence of a superflow, it is possible for quasiparticles to scatter, due solely to the antiperiodic boundary conditions. The effect of this Berry phase is considered in a companion paper.

Second, recall that the quasiparticle excitation spectrum for a $d$-wave superconductor is that of an anisotropic Dirac cone with two characteristic velocities, $v_{f}$ and $v_{2}$, which can be quite different. This anisotropy enters the BdG equation and current functional in the form of the parameter $\alpha=v_{2}/v_{f}$. The fact that $\alpha$ is not equal to one has the effect of complicating the form of the BdG equation in polar coordinates. Thus, in order to make this partial differential equation separable, we should scale out the anisotropy. This can be accomplished by scaling the $y$-coordinate by a factor of $\alpha$. In the scaled coordinates, the elliptical gap nodes become circular and
we obtain the isotropic Dirac equation. Unfortunately, this scaling also has a negative consequence. Whereas the vortex was originally circular in real space, the scaling makes it elliptical. Thus, in the scaled coordinates, the superfluid momentum (and the vortex boundary) become much more complicated. While such complications can be dealt with (via a significantly more involved computation), they were not considered in this investigation. Possible consequences of anisotropy have been studied in Ref. \onlinecite{Melnikov},  where it was noted that  there are no drastic experimental signatures due to the anisotropy of the Dirac cone. Thus, for simplicity, rather than scaling out the anisotropy, we consider the more straightforward case where the gap nodes are isotropic and $v_{f}=v_{2}$. For this isotropic case, both the gap nodes and the vortex are circular and we can separate the BdG equation in polar coordinates. However, since $v_{f}$ exceeds $v_{2}$ by a factor of 10 to 20 in the cuprates \cite{chiao}, this is clearly an approximation.
As a result, we expect only qualitative agreement with reality.  That this is the case was demonstrated by Durst, Vishwanath, and Lee\cite{adamprl} who used this approximation to provide a clear, though qualitative, explanation of the thermal Hall conductivity measurements of Ong and co-workers\cite{ong}

Finally, recall that the BdG Hamiltonian can be expressed as the sum of a linearized (Dirac) part, $H_{D}$, and a quadratic (curvature) part, $H_{C}$. We see that $H_{C}$ is small compared to $H_{D}$ as long as we are sufficiently far from the vortex center ($r > 1/p_{F}$). Hence we must cutoff our model at the scale of the vortex core ($\xi \sim 10/p_{F}$). As our final assumption, we select a reasonable core size and model the vortex core as a region with vanishing superflow. We now have a well-defined scattering problem, which is solved considering the linearized Hamiltonian. Since we shall only consider the isotropic case from this point forward, the Hamiltonian becomes
\begin{equation}
H_{D} = v_{f} \left[ \tau_{3} p_{x} + \tau_{1} p_{y} + P_{sx} \right]
\label{eq:isoHD}
\end{equation}

the quasiparticle energy is
\begin{equation}
E= v_{f} \sqrt{k_{x}^{2} + k_{y}^{2}} = v_{f} k
\label{eq:isoE}
\end{equation}
and the quasiparticle current functional takes the form
\begin{equation}
{\bf j}_{D} = v_{f} \Phi^{\dagger} (\tau_{3} \hat{\bf x}
+ \tau_{1} \hat{\bf y}) \Phi
\label{eq:isoJD}
\end{equation}

Note that the effective potential induced by the superflow is smooth on the scale of $1/p_F$.  The Fourier components for large-momentum inter-node scattering are therefore small.  Hence, we expect that the contribution to the scattering cross section of inter-node scattering will be subdominant to that of intra-node scattering.  Thus, for simplicity, we include only intra-node scattering in what follows.

\section{Angular Momentum Eigenstates}
\label{sec:angmom}
Consider the linearized Hamiltonian, $H_{D}$, in the isotropic
limit, Eq.~(\ref{eq:isoHD}). Note that for $P_{s}=0$, this is just the
Dirac Hamiltonian for massless spin-1/2 fermions in two dimensions
\begin{equation}
H_{D}^{0} = v_{f} [ \tau_{3} p_{x} + \tau_{1} p_{y}]
= v_{f} [ {\bf \alpha} \cdot {\bf p} + \beta m ] \;\;\;\;\;\;\;\;
m=0 \;\;\;\;\;\; \beta=\gamma^{0} \;\;\;\;\;\;
\alpha^{i} = \gamma^{0} \gamma^{i}
\label{eq:DiracHam}
\end{equation}
with the $\gamma$-matrix representation
\begin{equation}
\gamma^{0} = \tau_{2} \;\;\;\;\;\;
\gamma^{1} = i\tau_{1} \;\;\;\;\;\;
\gamma^{2} = -i\tau_{3} \;\;\;\;\;\; \rightarrow \;\;\;\;\;\;
\beta = \tau_{2} \;\;\;\;\;\;
\alpha^{1} = \tau_{3} \;\;\;\;\;\;
\alpha^{2} = \tau_{1}
\label{eq:gammamatrix}
\end{equation}
For the physical Dirac equation we know that the total angular momentum operator takes the form, $J = L + \frac{1}{2} \Sigma$ where $\Sigma = \frac{i}{2} \epsilon_{3jk} \gamma^{j} \gamma^{k}$. Therefore, in the above $\gamma$-matrix representation
\begin{equation}
J = L + \frac{\Sigma}{2} = -i \frac{\partial}{\partial \phi}
+ \frac{\tau_{2}}{2} .
\label{eq:angmomJ}
\end{equation}
Evaluating the commutator with $H_{D}^{0}$ we find (as expected)
\begin{equation}
\left[ J,H_{D}^{0} \right]
= -\tau_{2} H_{D}^{0} + \tau_{2} H_{D}^{0} = 0 .
\label{eq:JcomH}
\end{equation}
Therefore, there exists a complete set of simultaneous eigenstates of
$H_{D}^{0}$ and $J$.

The eigenstates of $L=-i\frac{\partial}{\partial \phi}$ take the form
\begin{equation}
L \Psi = \ell \Psi \;\;\;\;\;\; \rightarrow \;\;\;\;\;\;
\Psi = e^{i \ell \phi}
\left( \begin{array}{c} a(r) \\ b(r) \end{array} \right)
\label{eq:Leigen}
\end{equation}
and the eigenstates of $\Sigma = \tau_{2}$ take the form
\begin{equation}
\Sigma \Psi = \lambda \Psi \;\;\;\;\;\; \rightarrow \;\;\;\;\;\;
\lambda = \pm 1 \;\;\;\;\;\;
\Psi_{+} = c(r,\phi)
\left( \begin{array}{c} 1 \\ i \end{array} \right) \;\;\;\;\;\;
\Psi_{-} = d(r,\phi)
\left( \begin{array}{c} 1 \\ -i \end{array} \right)
\label{eq:tau2eigen}
\end{equation}
Therefore, the simultaneous eigenstates of $J$ and $H_{D}^{0}$ have the form
\begin{equation}
J \Phi_{n} = (n + 1/2) \Phi_{n} \;\;\;\;\;\;
\rightarrow \;\;\;\;\;\;
\Phi_{n} = f_{n}(r) e^{in\phi}
\left( \begin{array}{c} 1 \\ i \end{array} \right)
+ g_{n}(r) e^{i(n+1)\phi}
\left( \begin{array}{c} 1 \\ -i \end{array} \right)
\label{eq:Jeigen}
\end{equation}
where the radial functions, $f_{n}(r)$ and $g_{n}(r)$, are determined from the solution of a pair of coupled radial equations. Since we have neglected the Berry phase contribution and adopted periodic boundary conditions for this case, the requirement of single-valued wave functions demands that $n$ is an integer.

For $P_{s} \neq 0$, $H_{D}$ looks like the massless Dirac equation in the presence of an effective scalar potential,
$V = v_{f}P_{sx} = -v_{f}P_{s}(r) \sin \phi$. Since the effective potential is non-central, it mixes angular momentum eigenstates. (See Refs.~\onlinecite{mittleman} and \onlinecite{altshuler} for a discussion of electron scattering from a similar (dipole) potential.) While the general solution can still be expressed as a linear combination of angular momentum eigenstates
\begin{equation}
\Phi({\bf r}) = \sum_{n} \left[
f_{n}(r) e^{in\phi}
\left( \begin{array}{c} 1 \\ i \end{array} \right)
+ g_{n}(r) e^{i(n+1)\phi}
\left( \begin{array}{c} 1 \\ -i \end{array} \right) \right]
\label{eq:Phiform}
\end{equation}
the radial equations for different $n$ are now all coupled together. Nonetheless, general solutions can still be written in the form of Eq.~(\ref{eq:Phiform}). In the following section, we apply $H_{D}$ to a wave function of this form and proceed to determine the resulting radial equations.

\section{Radial Equations}
\label{sec:radeq}
We shall now plug the general form of our wave function, Eq.~(\ref{eq:Phiform}), into the Bo\-go\-liu\-bov-de Gennes equation. For this purpose, it is useful to write the BdG Hamiltonian, $H=H_{D}$, in polar coordinates. Doing so, we find
\begin{equation}
H_{D} = -iv_{f} \left[
({\bf \tau} \cdot \hat{\bf r}) \frac{\partial}{\partial r}
+ \frac{({\bf \tau} \cdot \hat{\bf r})}{r} i\tau_{2}
\frac{\partial}{\partial \phi} - P_{s} i \sin\phi \right]
\label{eq:radHD}
\end{equation}

where $({\bf \tau} \cdot \hat{\bf r}) \equiv \tau_{3} \cos\phi
+ \tau_{1} \sin\phi$ and we note that
$({\bf \tau} \cdot \hat{\bf r})^{2}=1$.
It is also useful to write the wave function as
\begin{equation}
\Phi({\bf r}) = \sum_{n} \left[
f_{n}(r) \chi_{n}^{+} + g_{n}(r) \chi_{n}^{-} \right]
\label{eq:Phichi}
\end{equation}
where we have defined
\begin{equation}
\chi_{n}^{+} \equiv e^{in\phi}
\left( \begin{array}{c} 1 \\ i \end{array} \right) \;\;\;\;\;\;\;\;
\chi_{n}^{-} \equiv e^{i(n+1)\phi}
\left( \begin{array}{c} 1 \\ -i \end{array} \right) .
\label{eq:chidef}
\end{equation}
Note that $({\bf \tau} \cdot \hat{\bf r})$ transforms $\chi_{n}^{+}$
into $\chi_{n}^{-}$ and vice versa,
\begin{equation}
({\bf \tau} \cdot \hat{\bf r}) \chi_{n}^{+} = \chi_{n}^{-} \;\;\;\;\;\;\;\;
({\bf \tau} \cdot \hat{\bf r}) \chi_{n}^{-} = \chi_{n}^{+} ,
\label{eq:chi1}
\end{equation}
and also that cosines and sines shift the $\chi$'s up and down
in angular momentum,
\begin{equation}
\cos\phi \chi_{n}^{\pm} = \frac{ \chi_{n+1}^{\pm} + \chi_{n-1}^{\pm} }{2}
\;\;\;\;\;\;\;\;
\sin\phi \chi_{n}^{\pm} = \frac{ \chi_{n+1}^{\pm} - \chi_{n-1}^{\pm} }{2i} .
\label{eq:chi2}
\end{equation}
Making use of these relations, we see that both $H\Phi$ and $E\Phi$ can be written as sums over $\chi_{n}^{\pm}$ weighted by coefficients that are functions only of the radial coordinate. Equating coefficients of $\chi_{n}^{\pm}$ on both sides of the BdG equation and dividing out an overall factor of $-iv_{f}k$ yields a set of fully coupled differential equations for
$f_{n}$ and $g_{n}$,
\begin{eqnarray}
\frac{\partial f_{n}}{\partial \rho} - \left[ \frac{n}{\rho} f_{n}
+ i g_{n} + \frac{P_{s}}{2} (g_{n-1} - g_{n+1}) \right] =0
\label{eq:feq}
\end{eqnarray}
\begin{eqnarray}
\frac{\partial g_{n}}{\partial \rho} - \left[ -\frac{n+1}{\rho} g_{n}
+ i f_{n} + \frac{P_{s}}{2} (f_{n-1} - f_{n+1}) \right] =0
\label{eq:geq}
\end{eqnarray}
where we have defined a dimensionless radial coordinate, $\rho \equiv kr$, and a dimensionless superfluid momentum,
$P_{s}(\rho) \equiv (1/\rho - \rho/(kR)^2)/2$. In this manner, the solution of the BdG equation is reduced to the that of a system of coupled ordinary differential equations.

\section{Outside the Vortex}
\label{sec:outside}
According to our model of the single vortex, the superfluid momentum vanishes at a distance, $R$, from the origin. Thus, for $\rho > kR$, we sit outside the vortex. Here $P_{s}=0$ and the quasiparticles are free. Since the superfluid momentum (Eq.~\ref{eq:ps}) includes the effects of both the order parameter phase gradient and the magnetic field, quasiparticles outside the vortex are subject to neither. Thus, for $\rho > kR$, we consider quasiparticles subject only to the free Dirac Hamiltonian, $H_{D}^{0}$, and with current defined via the linearized current functional, ${\bf j}_{D}[\Phi]$. In this regime, it is possible to obtain explicit solutions to the BdG equation, define incident and scattered wave functions, and, by constructing these incident and scattered waves from the free basis functions, write down an expression for the scattering cross section in terms of the coefficients of the basis functions. Then we need only match solutions with those inside the vortex to obtain the cross section.

\subsection{Free Solutions}
For $P_{s}=0$ , the radial equations, (\ref{eq:feq}) and (\ref{eq:geq}), take the dramatically simpler form
\begin{equation}
\left( \frac{\partial}{\partial \rho} - \frac{n}{\rho} \right)
f_{n} = i g_{n} \;\;\;\;\;\;\;\;
\left( \frac{\partial}{\partial \rho} + \frac{n+1}{\rho} \right)
g_{n} = i f_{n} .
\label{eq:freerad}
\end{equation}
Note in particular that, while $f_{n}$ is coupled to $g_{n}$, the equations for functions of different $n$ are independent. Eliminating $g_{n}$ from the equations above yields
\begin{equation}
\left[ \frac{\partial^{2}}{\partial \rho^{2}}
+ \frac{1}{\rho} \frac{\partial}{\partial \rho}
+ \left( 1- \frac{n^{2}}{\rho^{2}} \right) \right] f_{n} = 0
\label{eq:besseldef}
\end{equation}
which is the defining equation for the Bessel functions. Therefore,
\begin{equation}
f_{n}(\rho) = A_{n} J_{n}(\rho) + B_{n} Y_{n}(\rho)
\label{eq:freef}
\end{equation}
where $A_{n}$ and $B_{n}$ are complex constants and $J_{n}(\rho)$ and $Y_{n}(\rho)$ are Bessel functions of the first and second kind. Substituting back for $g_{n}$ then yields
\begin{equation}
g_{n}(\rho) = iA_{n} J_{n+1}(\rho) + iB_{n} Y_{n+1}(\rho)
\label{eq:freeg}
\end{equation}
where we have used the Bessel function identity
$(\partial/\partial\rho - \nu/\rho) Z_{\nu} = -Z_{\nu+1}$.
Hence, we can write down the free wave function
\begin{equation}
\Phi({\bf r}) = \sum_{n} \left[
(A_{n} J_{n} + B_{n} Y_{n}) e^{in\phi}
\left( \begin{array}{c} 1 \\ i \end{array} \right)
+ i(A_{n} J_{n+1} + B_{n} Y_{n+1}) e^{i(n+1)\phi}
\left( \begin{array}{c} 1 \\ -i \end{array} \right) \right]
\label{eq:freePhi}
\end{equation}
where the coefficients shall remain undetermined until we match with solutions inside the vortex.

\subsection{\label{sub:planewave}Incident and Scattered Waves}

We wish to obtain the cross section for a plane wave, with quasiparticle current in the incident direction, scattering off a vortex as a radial wave, with quasiparticle current in the scattered direction. If the incident momentum is ${\bf k} = (k,\theta)$ and the final momentum is ${\bf k}^{\prime} = (k,\phi)$, then the incident direction is the direction of the group velocity at momentum ${\bf k}$ and the scattered direction is the direction of the group velocity
at momentum ${\bf k}^{\prime}$. For general, anisotropic nodes, the group velocity need not be parallel to the momentum. However, for the isotropic case that we consider
\begin{equation}
{\bf v}_{G}({\bf k}) = \frac{\partial E_{k}}{\partial {\bf k}}
= v_{f} \frac{\epsilon_{k}}{E_{k}} \hat{\bf x}
+ v_{2} \frac{\Delta_{k}}{E_{k}} \hat{\bf y}
= v_{f} \left( \cos\theta \hat{\bf x} + \sin\theta \hat{\bf y} \right)
= v_{f} \hat{\bf k}
\label{eq:groupvel}
\end{equation}
and the group velocity and momentum are parallel. Therefore, if $\Phi_{i}$ denotes the incident wave function and $\Phi_{s}$ denotes the scattered wave function, then we require
\begin{equation}
{\bf j}_{D}[\Phi_{i}] \sim
\left( \cos\theta \hat{\bf x} + \sin\theta \hat{\bf y} \right)
\sim \hat{\bf k} \;\;\;\;\;\;\;\;
{\bf j}_{D}[\Phi_{s}] \sim
\left( \cos\phi \hat{\bf x} + \sin\phi \hat{\bf y} \right)
\sim \hat{\bf k}^{\prime} \sim \hat{\bf r} .
\label{eq:JincJscat}
\end{equation}

Recall that outside the vortex, quasiparticles are subject to neither an order parameter phase gradient nor a magnetic field.  Thus, the incident wave function is a plane wave.  This is consistent with the well-known results of Franz and Tesanovic (Franz and Tesanovic, PRL 84, 554 (2000)) who showed that the low-energy quasiparticle states of a d-wave superconductor in the vortex state are Bloch waves of massless Dirac fermions rather than Landau Levels.  (For a discussion of the analyses that led to this important result, the reader is referred to Refs.~\onlinecite{gorkovSchrieffer}, \onlinecite{kopninbook}, \onlinecite{kopninvinokur}, and \onlinecite{franztesano}.). Inspection of the form of the current functional,
${\bf j}_{D} = v_{f} \Phi^{\dagger}
(\tau_{3}\hat{\bf x} + \tau_{1}\hat{\bf y}) \Phi$
reveals that the appropriate incident plane wave is
\begin{equation}
\Phi_{i}({\bf r}) = e^{i{\bf k} \cdot {\bf r}}
\left( \begin{array}{c} \cos \frac{\theta}{2} \\
\sin \frac{\theta}{2} \end{array} \right)
\label{eq:Phii}
\end{equation}
since
\begin{equation}
{\bf j}_{D}[\Phi_{i}] = v_{f} \left[
\left( \cos^{2} \frac{\theta}{2} - \sin^{2} \frac{\theta}{2} \right)
\hat{\bf x} + \left( 2 \sin \frac{\theta}{2} \cos \frac{\theta}{2} \right)
\hat{\bf y} \right] = v_{f} \hat{\bf k} .
\label{eq:jinc}
\end{equation}

Note also that this form solves the BdG equation, as it must in the absence of the vortex. The appropriate scattered radial wave is then
\begin{equation}
\Phi_{s}({\bf r}) = e^{i\frac{\phi}{2}} f(\phi-\theta) \frac{e^{ikr}}{\sqrt{r}}
\left( \begin{array}{c} \cos \frac{\phi}{2} \\
\sin \frac{\phi}{2} \end{array} \right)
\label{eq:Phis}
\end{equation}
since
\begin{equation}
{\bf j}_{D}[\Phi_{s}] = v_{f} \frac{|f|^{2}}{r} \left[
\left( \cos^{2} \frac{\phi}{2} - \sin^{2} \frac{\phi}{2} \right)
\hat{\bf x} + \left( 2 \sin \frac{\phi}{2} \cos \frac{\phi}{2} \right)
\hat{\bf y} \right] = v_{f} \frac{|f|^{2}}{r} \hat{\bf r} .
\label{eq:jscat}
\end{equation}
Here $f(\phi-\theta)$ is the scattering amplitude and the $e^{i\phi/2}$ prefactor has been added to make the
wave function single-valued.

\subsection{Constructing the Cross Section}
\label{ssec:construct}
We shall now construct, from our free solution basis functions, an asymptotic wave function containing the correct incident and scattered waves. The factor in front of the scattered wave will then be our scattering amplitude.

We begin with the free wave function obtained in Eq.~({\ref{eq:freePhi}). Note that in the asymptotic limit, the integer-index Bessel functions can be expressed as
\begin{equation}
J_{n}(\rho) = \eta_{n} \sqrt{\frac{2}{\pi\rho}} \cos
(\rho - \pi/4 - |n|\pi/2) \;\;\;\;\;\;\;\;
Y_{n}(\rho) = \eta_{n} \sqrt{\frac{2}{\pi\rho}} \sin
(\rho - \pi/4 - |n|\pi/2)
\label{eq:asyBessel}
\end{equation}
where $\eta_{n}=1$ for $n \geq 0$ and $\eta_{n}=(-1)^{n}$
for $n < 0$.
Therefore, if we shift $n \rightarrow n-1$ in the second term of Eq.~({\ref{eq:freePhi}), plug in the asymptotic forms,
and decompose the sines and cosines into exponentials, we obtain
\begin{eqnarray}
\lefteqn{\Phi = \sum_{n} e^{in\phi} \sqrt{\frac{2}{\pi\rho}} \frac{\eta_{n}}{2}
\left[ \left( (A_{n}-iB_{n}) e^{i(\rho - \frac{\pi}{4} - \frac{\pi}{2}|n|)}
+ (A_{n}+iB_{n}) e^{-i(\rho - \frac{\pi}{4} - \frac{\pi}{2}|n|)}
\right) \left(\begin{array}{c} 1 \\ i \end{array}\right) \right.} \nonumber \\
&& \left.
+ \left( i(A_{n-1}-iB_{n-1}) e^{i(\rho - \frac{\pi}{4} - \frac{\pi}{2}|n|)}
+ i(A_{n-1}+iB_{n-1}) e^{-i(\rho - \frac{\pi}{4} - \frac{\pi}{2}|n|)}
\right) \left(\begin{array}{c} 1 \\ -i \end{array}\right) \right] .
\nonumber \\ &&
\label{eq:Phiasy1}
\end{eqnarray}
We can replace our two complex constants, $A_{n}$ and $B_{n}$, with two new complex constants, $a_{n}$ and $b_{n}$, by defining
\begin{equation}
A_{n}-iB_{n} \equiv i^{n} e^{-i(n+\frac{1}{2})\theta}
\left( 1/2 + e^{i\frac{\pi}{4}} b_{n} \right)
\label{eq:AmiB}
\end{equation}
\begin{equation}
A_{n}+iB_{n} \equiv i^{n} e^{-i(n+\frac{1}{2})\theta}
\left( 1/2 + e^{i\frac{\pi}{4}} (-1)^{n} a_{n} \right) .
\label{eq:ApiB}
\end{equation}
Making use of these definitions, noting that $\eta_{n} i^{n} = i^{|n|}$, and reorganizing terms, we find
\begin{eqnarray}
\lefteqn{\Phi = \sum_{n} i^{n} e^{in(\phi-\theta)} \eta_{n}
\sqrt{\frac{2}{\pi\rho}} \cos (\rho - \pi/4 -|n|\pi/2) \frac{1}{2}
\left[ e^{-i\frac{\theta}{2}}
\left(\begin{array}{c} 1 \\ i \end{array}\right)
+ e^{i\frac{\theta}{2}}
\left(\begin{array}{c} 1 \\ -i \end{array}\right)
\right]} \nonumber \\
&& \!\!\!\!\!\!\!\!
+ \sum_{n} e^{in(\phi-\theta)} \sqrt{\frac{2}{\pi\rho}} \frac{1}{2}
\left[ \left( b_{n} e^{i\rho} + a_{n} e^{-i\rho} \right)
e^{-i\frac{\theta}{2}}
\left(\begin{array}{c} 1 \\ i \end{array}\right)
+ \left( b_{n-1} e^{i\rho} - a_{n-1} e^{-i\rho} \right)
e^{i\frac{\theta}{2}}
\left(\begin{array}{c} 1 \\ -i \end{array}\right)
\right] \nonumber \\ &&
\label{eq:Phiasy2}
\end{eqnarray}
Noting the Bessel function expansion of a plane wave
\begin{equation}
e^{i\rho\cos(\phi-\theta)} = \sum_{n} i^{n} J_{n}(\rho) e^{in(\phi-\theta)}
= \sum_{n} i^{n} e^{in(\phi-\theta)} \eta_{n}
\sqrt{\frac{2}{\pi\rho}} \cos (\rho - \pi/4 -|n|\pi/2)
\label{eq:BesselPW}
\end{equation}
and shifting $n \rightarrow n+1$ in the final term
of Eq.~(\ref{eq:Phiasy2}) yields
\begin{equation}
\Phi = e^{i{\bf k} \cdot {\bf r}}
\left(\begin{array}{c} \cos\frac{\theta}{2} \\
\sin\frac{\theta}{2} \end{array}\right)
+ \sqrt{\frac{2}{\pi\rho}} e^{i(\phi-\theta)/2} \sum_{n} e^{in(\phi-\theta)}
\left[ b_{n} e^{i\rho}
\left(\begin{array}{c} \cos\frac{\phi}{2} \\
\sin\frac{\phi}{2} \end{array}\right)
+ a_{n} e^{-i\rho}
\left(\begin{array}{c} -i\sin\frac{\phi}{2} \\
i\cos\frac{\phi}{2} \end{array}\right) \right]
\label{eq:Phiasy3}
\end{equation}
Regrouping terms and defining $\varphi \equiv \phi - \theta$, this becomes
\begin{equation}
\Phi = e^{i{\bf k} \cdot {\bf r}}
\left(\begin{array}{c} \cos\frac{\theta}{2} \\
\sin\frac{\theta}{2} \end{array}\right)
+ e^{i\frac{\varphi}{2}} \left[ f(\varphi) \frac{e^{ikr}}{\sqrt{r}}
\left(\begin{array}{c} \cos\frac{\phi}{2} \\
\sin\frac{\phi}{2} \end{array}\right)
+ i g(\varphi) \frac{e^{-ikr}}{\sqrt{r}}
\left(\begin{array}{c} -\sin\frac{\phi}{2} \\
\cos\frac{\phi}{2} \end{array}\right) \right]
\label{eq:Phiasy4}
\end{equation}
where
\begin{equation}
f(\varphi) \equiv \sqrt{\frac{2}{\pi k}} \sum_{n} b_{n} e^{in\varphi}
\label{eq:fphidef}
\end{equation}
\begin{equation}
g(\varphi) \equiv \sqrt{\frac{2}{\pi k}} \sum_{n} a_{n} e^{in\varphi} .
\label{eq:gphidef}
\end{equation}
Since we have yet to restrict $a_{n}$ and $b_{n}$, this asymptotic wave function is still totally general. However, we have succeeded in rearranging it into a suggestive form. The three terms above are easily understood. The first is our incident plane wave, the second is the outgoing radial wave, and the third is an incoming radial wave. By construction, we require an incident plane wave and an outgoing radial wave. To realize this scenario, our asymptotic boundary
conditions require that there be no additional incoming wave. Thus, we require $g(\varphi)=0$ and must therefore set
$a_{n}=0$ for all $n$. In terms of our original constants, this restriction requires that
\begin{equation}
B_{n} = i \left( A_{n} - A_{n}^{0} \right) \;\;\;\;\;\;\;\;
A_{n}^{0} \equiv \mbox{$\frac{1}{2}$} i^{n} e^{-i(n + \frac{1}{2})\theta}
\label{eq:BnfromAn}
\end{equation}
and sets
\begin{equation}
b_{n} = e^{-i\frac{\pi}{4}} \left( \frac{A_{n}}{A_{n}^{0}} - 1 \right) .
\label{eq:bndef}
\end{equation}
Furthermore, we obtain a simple form for the asymptotic wave function
\begin{equation}
\Phi({\bf r}) = e^{i{\bf k} \cdot {\bf r}}
\left(\begin{array}{c} \cos\frac{\theta}{2} \\
\sin\frac{\theta}{2} \end{array}\right)
+ e^{i\frac{\varphi}{2}} f(\varphi) \frac{e^{ikr}}{\sqrt{r}}
\left(\begin{array}{c} \cos\frac{\phi}{2} \\
\sin\frac{\phi}{2} \end{array}\right)
\label{eq:asyPhi}
\end{equation}
where we recall that ${\bf k}=(k,\theta)$, ${\bf r}=(r,\phi)$,
and $\varphi=\phi-\theta$.

Applying the current functional, ${\bf j}_{D}$, to both the incident and scattered parts of this wave function yields
the incident and scattered current density
\begin{equation}
{\bf j}_{i} = {\bf j}_{D}[\Phi_{i}] = v_{f} \hat{\bf k}
\;\;\;\;\;\;\;\;
{\bf j}_{s} = {\bf j}_{D}[\Phi_{s}] =
v_{f} \frac{|f(\varphi)|^{2}}{r} \hat{\bf r} .
\label{eq:jijs}
\end{equation}
Then the differential cross section is
\begin{equation}
\frac{d\sigma}{d\varphi} =
\frac{{\bf j}_{s} \cdot {\bf r}}{|{\bf j}_{i}|} =
|f(\varphi)|^{2} .
\label{eq:dCSdef}
\end{equation}
Integrating over $\varphi$ yields the total cross section
\begin{equation}
\sigma = \int_{-\pi}^{\pi} \!d\varphi\, \frac{d\sigma}{d\varphi}
= \frac{4}{k} \sum_{n} |b_{n}|^{2} .
\label{eq:CSdef}
\end{equation}
By weighting with a factor of $(1-\cos\varphi)$, we obtain
the transport cross section
\begin{equation}
\sigma_{\parallel} = \int_{-\pi}^{\pi} \!d\varphi\, \frac{d\sigma}{d\varphi}
(1-\cos\varphi) = \frac{4}{k} \sum_{n}
b_{n} \left[ b_{n} - (b_{n+1}+b_{n-1})/2 \right]^{*}
\label{eq:transCSdef}
\end{equation}
and by weighting with a factor of $\sin\varphi$ we find the
the skew cross section
\begin{equation}
\sigma_{\perp} = \int_{-\pi}^{\pi} \!d\varphi\, \frac{d\sigma}{d\varphi}
\sin\varphi = \frac{2}{ik} \sum_{n}
b_{n} \left[ b_{n+1} - b_{n-1} \right]^{*} .
\label{eq:skewCSdef}
\end{equation}
All that remains is to calculate the coefficients $b_{n}$. These, of course, are determined by the details of the
quasiparticle scattering. Hence, we must now look inside the vortex.

\section{Inside the Vortex}
\label{sec:inside}
Inside the vortex, the situation is more complicated. For $\rho < kR$, the superfluid momentum is nonzero and takes the form, $P_{s} = (1/\rho - \rho/(kR)^{2})/2$. Therefore, even the linearized Hamiltonian, $H_{D}$, mixes angular momentum eigenstates. As discussed by Simon and Lee\cite{simon}, Ye\cite{Ye}, and Vishwanath\cite{ashwin}, since the linearized Hamiltonian is time-reversal invariant, the resulting skew cross section (after summing over all four nodes) is zero. We solve for the radial functions, $f_{n}(\rho)$ and $g_{n}(\rho)$. The radial equations look as follows:
\begin{equation}
\frac{\partial f_{n}}{\partial \rho} - \left[ \frac{n}{\rho} f_{n}
+ i g_{n} + \frac{P_{s}}{2} (g_{n-1} - g_{n+1}) \right] = 0
\label{eq:feq1st}
\end{equation}
\begin{equation}
\frac{\partial g_{n}}{\partial \rho} - \left[ -\frac{n+1}{\rho} g_{n}
+ i f_{n} + \frac{P_{s}}{2} (f_{n-1} - f_{n+1}) \right] = 0
\label{eq:geq1st}
\end{equation}

Since the radial equations are fully coupled by the superfluid momentum terms, the equations for all $n$ should be solved simultaneously via an infinite-dimensional matrix equation. By numerical necessity, we shall cut off the coupling at some large $n$ such that we consider a total of $N$ angular momentum eigenstates where $-\frac{N}{2} \leq n \leq \frac{N}{2}-1$. This is physically reasonable as we do not expect very large angular momenta to have a significant effect on the low energy physics. The result is a $2N \times 2N$ matrix differential equation
\begin{equation}
\frac{d {\bf z}}{d \rho} = {\bf M}(\rho) {\bf z}
\label{eq:mateq}
\end{equation}
where ${\bf z}(\rho)$ is a $2N$-component vector containing the $f_{n}$'s and the $g_{n}$'s, ${\bf M}(\rho)$ is
a $2N \times 2N$ matrix. Since only neighboring angular momenta are coupled, ${\bf M}(\rho)$ is a rather sparse matrix. Furthermore, due to the form of Eqs.~(\ref{eq:feq1st}) and (\ref{eq:geq1st}) and the simple $\rho$-dependence of the superfluid momentum, this matrix can be written as
\begin{equation}
{\bf M}(\rho) = {\bf B} \frac{1}{\rho} + {\bf A}_{0} + {\bf A}_{1} \rho
\label{eq:MfromBA}
\end{equation}
where ${\bf B}$, ${\bf A}_{0}$, and ${\bf A}_{1}$
are constant $2N \times 2N$ matrices.
In terms of this matrix notation, our procedure will be to solve the homogenous equation.

\subsection{Homogeneous Solutions via Method of Frobenius}
\label{sec:frob}
Consider the homogeneous equation
\begin{equation}
\frac{d {\bf z}}{d \rho} = {\bf M}(\rho) {\bf z}
\;\;\;\;\;\; \rightarrow \;\;\;\;\;\;
\rho \frac{d {\bf z}}{d \rho}
= \left( {\bf B} + {\bf A}_{0} \rho + {\bf A}_{1} \rho^{2} \right)
{\bf z}
\label{eq:homog}
\end{equation}
Since ${\bf M}(\rho)$ diverges as $1/\rho$ at the origin, the equation has a regular singular point at $\rho=0$.
Thus, its solution requires local analysis about the origin. In particular, we shall employ a matrix generalization of what is known as the Method of Frobenius\cite{bender}.

Consider a solution of the (Frobenius) form
\begin{equation}
{\bf z} = \sum_{m=0}^{\infty} {\bf a}_{m} \rho^{\alpha + m}
\label{eq:frobform}
\end{equation}
where $\alpha$ is a complex number and ${\bf a}_{m}$ is a vector coefficient for each integer $m$. Plugging this into our differential equation yields
\begin{equation}
\sum_{m=0}^{\infty} \left[ \left( {\bf B} - (\alpha+m) \right)
{\bf a}_{m} \rho^{\alpha+m} + {\bf A}_{0} {\bf a}_{m} \rho^{\alpha+m+1}
+ {\bf A}_{1} {\bf a}_{m} \rho^{\alpha+m+2} \right] = 0 .
\label{eq:frob1}
\end{equation}
Shifting $m \rightarrow m-1$ for the ${\bf A}_{0}$ term, shifting $m \rightarrow m-2$ for the ${\bf A}_{1}$ term,
and noting that the total coefficient of each power of $\rho$ must equal zero, we obtain the following equations:
\begin{equation}
\left( {\bf B} - \alpha \right) {\bf a}_{0} = 0
\label{eq:Beig}
\end{equation}
\begin{equation}
\left( {\bf B} - (\alpha+1) \right) {\bf a}_{1} = - {\bf A}_{0} {\bf a}_{0}
\label{eq:rec1}
\end{equation}
\begin{equation}
\left( {\bf B} - (\alpha+m) \right) {\bf a}_{m}
= - {\bf A}_{0} {\bf a}_{m-1} - {\bf A}_{1} {\bf a}_{m-2}
\;\;\;\;\;\;\;\; m=2,3,4,\ldots
\label{eq:rec2}
\end{equation}
Note that Eq.~(\ref{eq:Beig}) is just the eigenvalue equation for ${\bf B}$. It is solved if $\alpha$ is an eigenvalue of ${\bf B}$ and ${\bf a}_{0}$ is the corresponding eigenvector. This eigensystem has $2N$ solutions which we can label $\alpha_{k}$ and ${\bf a}_{0}^{k}$ for $k=1 \ldots 2N$. The other two equations, (\ref{eq:rec1}) and (\ref{eq:rec2}), define matrix recursion relations. If ${\bf B} - (\alpha_{k}+m){\bf \openone}$ is non-singular for $m=1,2,3,\ldots$, then we can solve these equations to obtain ${\bf a}_{m}^{k}$ given ${\bf a}_{0}^{k}$ and $\alpha_{k}$. This condition will always be satisfied for the largest eigenvalue of {\bf B} but will fail for any $\alpha_{k}$ for which $\alpha_{k}+m$ is also an eigenvalue of {\bf B}. In general, this circumstance would require additional solutions not of the Frobenius form. Fortunately, for $P_{s} \neq 0$, the eigenvalues of {\bf B} are separated by non-integer
numbers. Therefore, we can obtain $2N$ solutions of the Frobenius form
\begin{equation}
{\bf z}^{k} = \sum_{m=0}^{m_{\mathrm{max}}} {\bf a}_{m}^{k} \rho^{\alpha_{k} + m}
\;\;\;\;\;\;\;\; k=1 \ldots 2N
\label{eq:zksolns}
\end{equation}
and use the equations above to calculate the vector coefficients recursively. Due to the constraints of our numerics, we cutoff each series at $m=m_{\mathrm{max}}$. As such series represent expansions about the origin, the more terms we include the larger the $\rho$ at which the solutions will be valid. Since the differential equation contains no
singular points besides the origin, the solution can be pushed as far from the origin as necessary by including additional terms. Since our solutions need only be valid out to the edge of the vortex, $\rho=kR$, we shall cutoff
each series at the $m_{\mathrm{max}}$ for which the solutions are valid out to $kR$.

\subsection{Boundary Condition at Origin}
\label{ssec:bcorigin}
Although all $2N$ solutions, ${\bf z}^{k}$, satisfy the homogeneous equation (\ref{eq:homog}), we must reject solutions that show unphysical behavior at the origin. While it is permissible for the wave function or the current density to diverge at the origin, we must require that observable quantities like the total probability at the origin and the total current passing through the origin be well behaved. Since solutions of the Frobenius form yield contributions to the wave function of the form $\Phi(\rho \rightarrow 0) \sim \rho^{\alpha}$ for small $\rho$, we seek conditions on the allowed values of $\alpha$.

Consider a circle of radius $\epsilon \rightarrow 0$ about the origin. The total quasiparticle probability within the circle is
\begin{equation}
\mbox{Prob} = \int^{\epsilon} \!d^{2}\rho\,
\Phi^{\dagger}(\rho \rightarrow 0) \Phi(\rho \rightarrow 0)
\sim \int_{0}^{\epsilon} \!(\rho d\rho)\, \rho^{\alpha} \rho^{\alpha}
\sim \int_{0}^{\epsilon} \! d\rho \, \rho^{2\alpha+1} .
\label{eq:originprob}
\end{equation}
If the origin is to be just like any other point in space, then this probability must vanish as $\epsilon \rightarrow 0$. Hence, we require $\alpha > -1$. While this restricts the allowed solutions, it turns out that the
quasiparticle current provides a more restrictive condition.

Now consider a semicircle of radius $\epsilon \rightarrow 0$ about the origin, oriented about the $\hat{\bf \theta}$ direction. The total current passing though such a semicircle in the $\hat{\bf \theta}$ direction is
\begin{equation}
I_{\theta} = \int_{\theta-\pi/2}^{\theta+\pi/2}
\! (\epsilon d\phi) \, v_{f} \Phi^{\dagger}(\rho \rightarrow 0)
(\tau_{3} \hat{\bf x} + \tau_{1} \hat{\bf y}) \Phi(\rho \rightarrow 0)
\cdot \hat{\bf \theta}
\sim \epsilon \epsilon^{\alpha} \epsilon^{\alpha}
\sim \epsilon^{2\alpha+1} .
\label{eq:origincurrent}
\end{equation}
For the origin to be physical, this current must vanish as $\epsilon \rightarrow 0$. Hence, we require $\alpha \geq -1/2$.

For $P_{s} \neq 0$, the eigenvalues of ${\bf B}$ are always such that $N$ of them are strictly larger than $-1/2$ and
$N$ of them are strictly smaller than $-1/2$. Thus, in order to satisfy these boundary conditions at the origin, we shall keep only the $N$ solutions, ${\bf z}^{k}$, for which $\alpha_{k} > -1/2$. The full solution to the homogeneous equation is then
\begin{equation}
{\bf z}(\rho) = \sum_{k=1}^{N} c_{k} {\bf z}^{k}(\rho)
\label{eq:homosoln}
\end{equation}
where the $c_{k}$ are $N$ complex coefficients to be determined via solution matching at the edge of the vortex.

\section{Four-Node Cross Section}
\label{sec:4node}
For $\rho > kR$, we found $2N$ radial solutions (Bessel functions) and reduced the $2N$ coefficients ($A_{n}$,$B_{n}$) to $N$ coefficients with the condition that there be no incident radial wave in the asymptotic limit.
(This required $B_{n}=i(A_{n}-A_{n}^{0})$.) For $\rho < kR$, we found $2N$ radial solutions (Frobenius series) and reduced the $2N$ coefficients to $N$ coefficients with the condition that
the solutions be well-behaved at the origin. (This required $c_{k}=0$ for $\alpha_{k} < -\frac{1}{2}$.)
At $\rho=kR$, all $2N$ radial solutions must be continuous. (Note that since the Dirac equation is 1st order, only continuity is required, while for the 2nd order Schrodinger equation, differentiability would also be required.) Thus, we have $2N$ equations which can be solved for the remaining $2N$ coefficients by matching solutions at $\rho=kR$.
We solve the resulting $2N \times 2N$ matrix equation via LU decomposition to obtain $c_{k}$ for $1 \leq k \leq N$
and $A_{n}$ for $-\frac{N}{2} \leq n \leq \frac{N}{2}-1$.

As discussed in prior sections, solutions are matched to obtain to obtain the coefficients to get scattering cross sections. Recall that at the outset of this calculation, we shifted the origin of momentum space to the center of node 1. Thus, in the discussions that followed, we have been considering quasiparticles scattered from one state in the vicinity of node 1 to another state in the vicinity of node 1. The resulting cross section is therefore only the
cross section for these node-1 quasiparticles. However, given a quasiparticle current in any particular direction, quasiparticles from all four nodes will contribute equally. Thus to obtain the physical cross section, we must average over the cross sections for quasiparticles at each of the four nodes.

Our results for node 1 can be easily generalized to node $j=\{1,2,3,4\}$ by transforming coordinates to those appropriate to node $j$. In accordance with the $d$-wave structure of the gap, we can define a local coordinate
system at each of the four nodes with a $\hat{\bf k}_{1}$ axis pointing along the direction of increasing $\epsilon_{k}$ and a $\hat{\bf k}_{2}$ axis pointing along the direction of increasing $\Delta_{k}$. Note that while nodes 1 and 3 define right-handed coordinate systems, nodes 2 and 4 define left-handed coordinate systems. We can therefore transform from node 1 to node $j$ simply by rotating our incident and scattered angles ($\theta$ and $\phi$) and then changing the sign of these angles to account for the handedness of the local coordinate system.
\begin{equation}
\begin{array}{llll}
\mbox{Node 1:} & \theta_{1}=\theta
& \phi_{1}=\phi
& \varphi_{1} = \phi_{1}-\theta_{1} = \varphi \\
\mbox{Node 2:} & \theta_{2}=-(\theta-\mbox{$\frac{\pi}{2}$})
& \phi_{2}=-(\phi-\mbox{$\frac{\pi}{2}$})
& \varphi_{2} = \phi_{2}-\theta_{2} = -\varphi \\
\mbox{Node 3:} & \theta_{3}=\theta+\pi
& \phi_{3}=\phi+\pi
& \varphi_{3} = \phi_{3}-\theta_{3} = \varphi \\
\mbox{Node 4:} & \theta_{4}=-(\theta+\mbox{$\frac{\pi}{2}$})
& \phi_{4}=-(\phi+\mbox{$\frac{\pi}{2}$})
& \varphi_{4} = \phi_{4}-\theta_{4} = -\varphi
\end{array}
\label{eq:angles}
\end{equation}
Thus, to obtain results for quasiparticles about node j, we need only input each $\theta_{j}$ and take the output
as a function of $(-1)^{j+1} \varphi$. Then the physical cross sections are
\begin{equation}
\frac{d\sigma}{d\varphi} = \frac{1}{4}\sum_{j=1}^{4}
\left( \frac{d\sigma}{d\varphi} \right)_{j}
\;\;\;\;\;\;
\sigma_{\parallel} = \frac{1}{4}\sum_{j=1}^{4} \sigma_{\parallel}^{j}
\label{eq:4nodeaverage}
\end{equation}

Note that we do not account for quasiparticles that are scattered from one node to another. However, since the
effective potential (induced by the superflow) is smooth on the scale of $1/p_{F}$, the Fourier components
for large-momentum inter-node scattering will be small. Hence, it is reasonable to neglect such contributions.

\section{Results}
\label{sec:vortexresults}
The procedure outlined in the preceding sections was implemented numerically for a range of intervortex distances,
$kR=\gamma T/\sqrt{H}$, from 0.5 to 15. The larger the $kR$, the more angular momentum eigenstates that contribute to the vortex scattering cross section, and the more runtime-intensive the computation. For $kR=15$, we
set $N=46$ and considered total angular momenta, $n+1/2$, ranging from -22.5 to 22.5. Calculating solutions to the BdG equation both inside and outside of the vortex, the program matches solutions at $\rho=kR$ to obtain the coefficients, $b_{n}$, first for node 1 and then for nodes 2, 3, and 4. For each of the four nodes, Eqs.~(\ref{eq:fphidef}) and (\ref{eq:dCSdef}) are then used to calculate the differential cross section for quasiparticles in the vicinity of that node. The coefficients and differential cross sections for $kR=8$ are shown in Fig.~\ref{fig:dCSpolar1} (Left). Polar plots of the four-node differential cross sections for integer $kR$ from 1 to 15 are shown in Fig. \ref{fig:dCSpolar1} (Right). In all cases, $d\sigma/d\varphi$ has a two peak structure and vanishes for $\varphi=\pi$. For small $kR$, the peaks are centered about $\pm\pi/2$. As $kR$ increases, the magnitude of $d\sigma/d\varphi$ increases while the peaks sweep closer to the forward direction.
\begin{figure}
  \includegraphics[width=6cm]{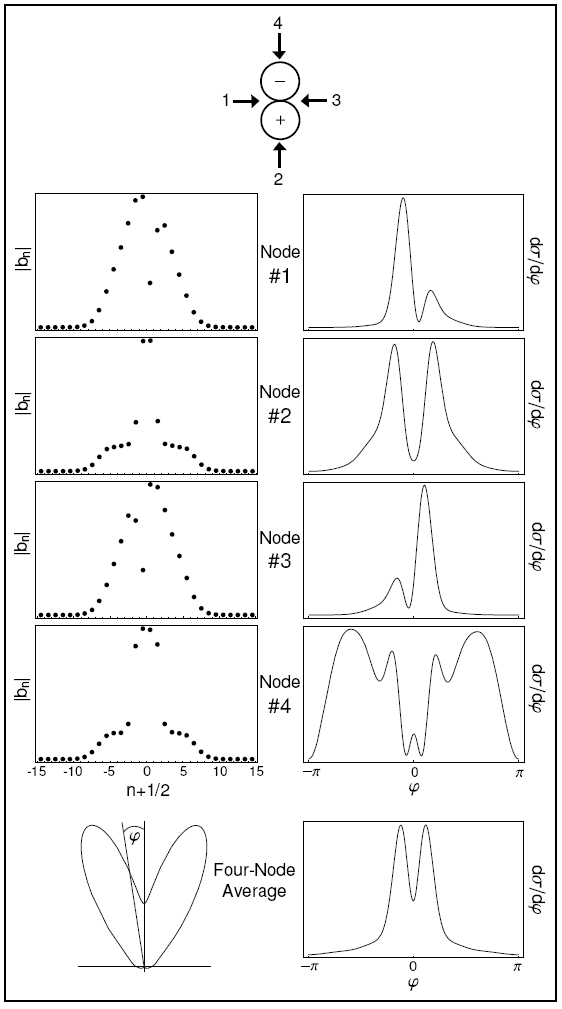}\qquad\qquad\includegraphics[width=6cm]{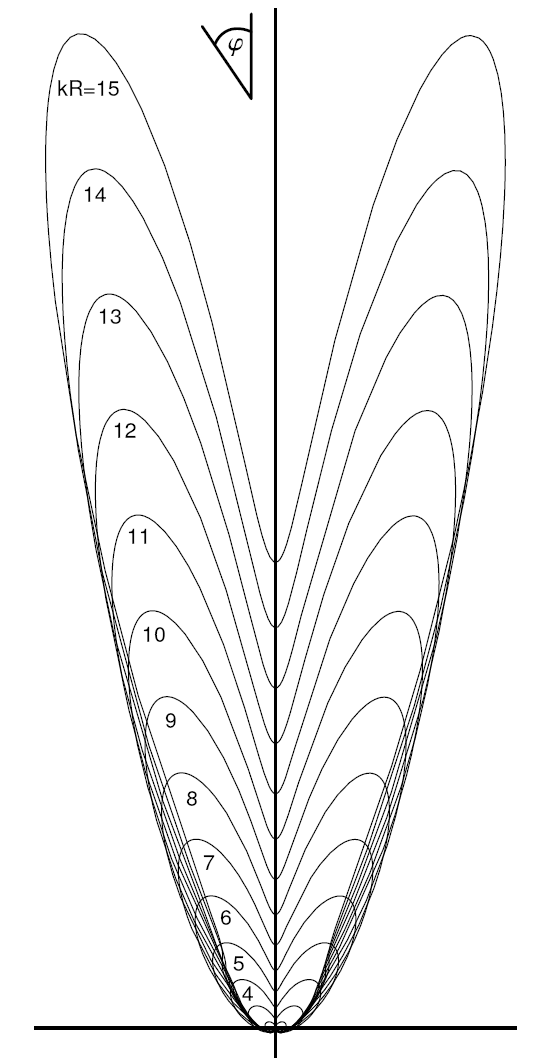}
\caption[Matched coefficients and differential cross sections for
quasiparticles about nodes 1, 2, 3, and 4.]
{(Left) Matched coefficients and differential cross sections for $kR=8$.
The coefficients, $b_{n}$, denote the contributions from eigenstates
with total angular momentum $n+1/2$. These coefficients and
the corresponding differential cross sections are plotted for
quasiparticles about nodes 1, 2, 3, and 4. The final row
contains the four-node average $d \sigma / d \varphi$ plotted
in both Cartesian and polar form. The schematic above indicates
the directions from which quasiparticles about each node approach the
non-central effective potential.
(Right) Fixed-scale polar plots of the four-node differential cross section for integer $kR$ from 1 to 15. Note that the cross section
magnitude grows with increasing $kR$.}
\label{fig:dCSpolar1}
\end{figure}



\begin{figure}
\includegraphics[scale=0.5]{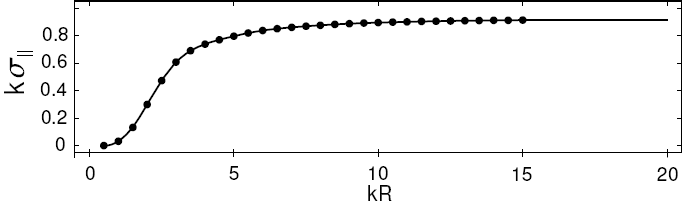}
\caption{Plot of $\sigma_{\|}$ (the solid line is an interpolation
of the numerical data in which the constant large-kR behavior is extrapolated to
larger kR)}
\label{fig:sigmasvp}
\end{figure}
The single vortex cross section discussed above was calculated to all orders in the linearized potential . However, to gain some physical insight regarding the calculated form of our results, it is instructive to consider the more transparent Born-limit calculation (valid to first order in the potential) whereby $d\sigma/d\varphi$
is proportional to the square of the Fourier transform of the scattering potential. For the linearized Hamiltonian, our effective potential is $V^{D}=-v_{f}(1/r-r/R^2)\theta(R-r)\sin\phi$ where the $\theta$-function imposes a cutoff at $r=R$. Given this input, the Born approximation yields an analytical expression for the differential cross section. Details of this calculation is given in Appendix~\ref{bornsc}. We find
\begin{equation}
\left. \frac{d\sigma}{d\varphi} \right|_{\mathrm{Born}} = \frac{\pi}{8k}
\frac{\cos^{2}(\varphi/2)}{1-\cos\varphi} \left( 1 -
\frac{J_{1}(2kR\sin(\varphi/2))}{kR\sin(\varphi/2)} \right)^{2}
\label{eq:borndCS}
\end{equation}
where $J_{1}(x)$ is the first order Bessel function of the first kind. This form comes about because the two-dimensional Fourier transform of $1/r$ yields $1/q$ where $q^{2}=|{\bf k}-{\bf k}^{\prime}|^{2}=2k(1-\cos\varphi)$. Hence
the square is proportional to $1/(1-\cos\varphi)$ and goes like $1/\varphi^{2}$ for small angles. The rest of the expression results from our cutoff at the vortex edge ($r=R$) and serves to suppress $d\sigma/d\varphi$ for $\varphi < 1/kR$. Integrating over angles therefore yields a total cross section that is linear in $kR$ for large $kR$.
Recall that the transport cross section includes an additional weighting factor of $(1-\cos\varphi)$. This precisely
cancels the leading functional dependence of $d\sigma/d\varphi$ and results in a transport cross section that approaches a constant for large $kR$. Our numerical results for $\sigma_{\parallel}$ (see Fig.~\ref{fig:sigmasvp}) are therefore quite logical. Thus, the Born approximation is sufficient to capture the qualitative features of our transport cross sections.
\section{Conclusions}
\label{conc}

In this paper, we considered the scattering of thermally-excited
quasiparticles from a single magnetic vortex in a $d$-wave
superconductor.  The scattering effect of the vortex can be divided
into two contributions, one due to the superflow circulating about the
vortex and another due to the Berry phase acquired by the
quasiparticle upon circling the vortex.  Both effects are present in
the Bogoliubov-de Gennes equation for a single vortex.  By applying a
singular gauge transformation, we have explicitly separated them, the
superflow resulting in an effective non-central potential and the
Berry phase generating antiperiodic boundary conditions for the
quasiparticle wave function.  In a separate paper \cite{gkd}, we have
studied the Berry phase contribution in the absence of the superflow
effect, essentially the Aharonov-Bohm scattering of a massless Dirac
fermion.  Here, by adopting periodic (rather than antiperiodic)
boundary conditions, we have considered the superflow contribution in
the absence of the Berry phase effect.  The resulting problem is that
of a massless Dirac fermion scattering from a non-central potential
(non-central because the gap node about which we linearize is shifted
away from the origin of momentum space).  We solved this problem by
calculating the eigenstates inside and outside the vortex, building an
incoming plane wave and an outgoing radial wave, and computing the
resulting differential cross section.  We plotted the size and shape
evolution of the differential cross section as a function of the ratio
of the magnetic length $R$ to the de Broglie wavelength $1/k$.
Intregrating, we computed the transport cross section and showed that
it saturates to a constant value for large $kR$.  This result was
previously used in Ref. \onlinecite{adamprl} to compute thermal transport coefficients.
 Here, we have provided the details of its derivation.  We have also
provided an alternate derivation, and a closed-form solution, valid
within the Born approximation.  These results, taken together with the
results of Ref.~\onlinecite{gkd}, shed significant light on the nature
of the scattering of massless Dirac quasiparticles from magnetic
vortices.  What remains is a complete analysis, taking into account
both the superflow and Berry phase contributions and the interference
between them.

\section{Acknowledgments}

We would all like to thank Sasha
Abanov, Patrick Lee, Zlatko Tesanovic, and Ashvin Vishwanath for very helpful discussions. This work was supported by the NSF under grant No. DMR-0605919. M.K. was also supported by the NSF under grant No. DMR-0906866. S.G. was also supported by the DOE under grant no. DE-FG02-09ER16052.

\appendix
\section{Born scattering for massless Dirac equation}
\label{bornsc}

In this appendix, we derive a closed-form expression for the superflow contribution to the single vortex cross section within the Born approximation\cite{sakurai}. We begin with a brief recap of Born scattering. The Hamiltonian can be written as
\begin{eqnarray}
H & = & H_{0}+V
\end{eqnarray}
where $H_{0}$ is the Kinetic energy operator
\begin{equation}
H_{0}=\frac{p^{2}}{2m}
\end{equation}
If we denote $\mid\phi>$ as the energy eigenket of $H_{0}$ then we have
\begin{equation}
H_{0}\mid\phi>=E\mid\phi>
\end{equation}
The Schrodinger equation we need to solve is
\begin{equation}
(H_{0}+V)\mid\psi>=E\mid\psi>
\end{equation}
The desired solution has the form
\begin{equation}
\mid\psi>=(E-H_{0})^{-1}V\mid\psi>+\mid\phi>
\end{equation}
In the position basis one could write it conveniently as
\begin{equation}
<x\mid\psi> = <x\mid\phi>+\int d^{2}x'<x\mid(E-H_{0})^{-1}\mid x'><x'\mid V\mid x>
\end{equation}
We will express the operator $(E-H_{0})^{-1}$ in the position basis as
\begin{equation}
<x\mid(E-H_{0})^{-1}\mid x'> = \int\int d^{2}p'd^{2}p''<x\mid\\ p'><p'\mid(E-H_{0})^{-1}\mid p''><p''\mid x'>
\end{equation}
where
\begin{equation}
<x\mid p> = \frac{1}{2\pi}e^{i\vec{p}\cdot\vec{x}}
\end{equation}
Also note that
\begin{eqnarray}
<p'\mid p>=\int d^{2}x'<p'\mid x><x\mid p>=\delta(\vec{p}-\vec{p'})
\end{eqnarray}
Now we apply this Born scattering method to the single vortex scattering of quasiparticles satisfying the Bogoliubov de-Gennes equation (\ref{eq:BdG}).
\begin{eqnarray}
H_{0} & = & v_{f}(\tau_{3}p_{x}+\tau_{1}p_{y})\quad \mbox{with} \quad E=v_{f}k\\
E-H_{0} & = & v_{f}(k-\tau_{3}p_{x}-\tau_{1}p_{y})=v_{f}\left(\begin{array}{cc}
k-p_{x} & -p_{y}\\
-p_{y} & k+p_{x}\end{array}\right)\\
(E-H_{0})^{-1} & = & \frac{1}{v_{f}}\frac{1}{k^{2}-p_{x}^{2}-p_{y}^{2}}\left(\begin{array}{cc}
k+p_{x} & p_{y}\\
p_{y} & k-p_{x}\end{array}\right)=\frac{1}{v_{f}}\frac{(k+\tau_{3}p_{x}+\tau_{1}p_{y})}{k^{2}-p^{2}}\\
<p'\mid(E-H_{0})^{-1}\mid p''> & = & \frac{1}{v_{f}}\frac{(k+\tau_{3}p_{x}+\tau_{1}p_{y})}{k^{2}-p^{2}}\delta(\vec{p}'-\vec{p''})\\
<x\mid(E-H_{0})^{-1}\mid x'> & =\frac{1}{v_{f}} & \int\frac{d^{2}p'}{(2\pi)^{2}}e^{i\ \vec{p'\cdot(}\vec{x}-\vec{x}')}\frac{(k+\tau_{3}p'_{x}+\tau_{1}p'_{y})}{k^{2}-p'^{2}}
\label{xEx}
\end{eqnarray}
 To evaluate the above integral Eq.~(\ref{xEx}) we define the angles specifying the incoming and outgoing momentum directions.
\begin{equation}
\vec{p'}=(p',\theta'),\qquad\vec{x}-\vec{x}'=(\mid\vec{x}-\vec{x}'\mid,\alpha),\ \ \ \varphi'=\alpha-\theta'
\label{angledef}
\end{equation}
Using the above definitions and plugging into Eq.~(\ref{xEx}) we have
\begin{eqnarray*}
<x\mid(E-H_{0})^{-1}\mid x'> & = & \frac{1}{v_{f}(2\pi)^{2}}\intop_{0}^{\infty}p'dp'\intop_{-\pi}^{\pi}d\varphi'e^{i p'\mid\vec{x}-\vec{x}'\mid\cos\varphi'}\frac{(k+p'(\tau_{3}\cos(\alpha-\varphi')+\tau_{1}\sin(\alpha-\varphi')))}{k^{2}-p'^{2}}\\
 & = & \frac{1}{v_{f}(2\pi)^{2}}\intop_{0}^{\infty}dp'\intop_{-\pi}^{\pi}d\varphi'e^{i p'\mid\vec{x}-\vec{x}'\mid\cos\varphi'}\frac{p'(k+p'\cos\varphi'(\tau_{3}\cos(\alpha)+\tau_{1}\sin(\alpha))}{k^{2}-p'^{2}}\\
 & = & \frac{1}{v_{f}(2\pi)^{2}}\intop_{0}^{\infty}dp'\frac{p'}{k^{2}-p'^{2}}((kI_{1}+p'M_{\alpha}I_{2})
\label{xEx1}
\end{eqnarray*}
where $I_{1}=\intop_{-\pi}^{\pi}d\varphi'e^{ia\cos\varphi'}=2\pi J_{0}(a)$ with $a=p'\mid\vec{x}-\vec{x}'\mid$
($J_{n}$ being the Bessels functions of first kind) and $M_{\alpha}=\tau_{3}\cos(\alpha)+\tau_{1}\sin(\alpha)$. The expression for $I_{2}$ is given by
\begin{eqnarray}
I_{2} & = & \intop_{-\pi}^{\pi}d\varphi'e^{ia\cos\varphi'}\cos\varphi'\\
 & = & -i\frac{\partial}{\partial a}\intop_{-\pi}^{\pi}d\varphi'e^{ia\cos\varphi'}=-i2\pi\frac{\partial}{\partial a}J_{0}(a)\\
 & = & 2\pi iJ_{1}(a)
\end{eqnarray}
This enables us to easily rewrite Eq.~(\ref{xEx1}) as
\begin{equation}
<x\mid(E-H_{0})^{-1}\mid x'>=\frac{1}{v_{f}(2\pi)}\intop_{0}^{\infty}dp'\frac{p'}{k^{2}-p'^{2}}[(kJ_{0}(p'\mid\vec{x}-\vec{x}'\mid)\ +ip'M_{\alpha}J_{1}(p'\mid\vec{x}-\vec{x}'\mid)]\end{equation}
We redefine the variables as $u=p'/k$ , $du=dp'/k$, and $z=k\mid\vec{x}-\vec{x}'\mid$. We also regularize the inverse operator by introducing $k\rightarrow k+i\epsilon$ with $\epsilon\rightarrow0$. The result is
\begin{eqnarray}
<x\mid(E-H_{0}+i\epsilon)^{-1}\mid x'> & = & \frac{k}{v_{f}(2\pi)}\intop_{0}^{\infty}du\frac{u}{1-u^{2}+i\epsilon}[J_{0}(zu)\ +iu\ M_{\alpha}J_{1}(z u)]\label{eq:regularinvpos}\\
 & = & \frac{-k}{v_{f}(2\pi)}\intop_{0}^{\infty}du\frac{u}{u^{2}-1-i\epsilon}[J_{0}(zu)\ +iu\ M_{\alpha}J_{1}(z u)]\label{eq:regularisedinvpos}\end{eqnarray}
We remind the reader of the following useful identities:
\begin{eqnarray}
\intop_{0}^{\infty}du\frac{u^{\nu+1}J_{\nu}(zu)}{u^{2}+b^{2}} & = & b^{\nu}K_{\nu}(b\ z)\quad\quad for\ z>0,\ Re[b]>0\ ,-1<Re[\nu]<\frac{3}{2}\\
K_{\nu}(z) & = & i\frac{\pi}{2}e^{i\frac{\pi}{2}\nu}H_{\nu}(z\ e^{i\frac{\pi}{2}\nu})\quad for\ -\pi<Arg(z)\leq\frac{\pi}{2}
\end{eqnarray}
where $H_{\nu}(z\ )$ are the Hankel functions of first kind.
Using the above identities in the integrals in Eq.~(\ref{eq:regularisedinvpos}) we obtain in the limit $\epsilon\rightarrow0$
\begin{eqnarray}
\intop_{0}^{\infty}du\frac{uJ_{\nu}(zu)}{u^{2}+(\epsilon-i)^{2}} & = & K_{\nu}(-i\ z)=i\frac{\pi}{2}H_{0}(z)\\
\intop_{0}^{\infty}du\frac{u^{2}J_{\nu}(zu)}{u^{2}+(\epsilon-i)^{2}} & = & i\frac{\pi}{2}H_{1}(z)
\end{eqnarray}
Hence, Eq.~(\ref{eq:regularisedinvpos}) reduces to
\begin{eqnarray}
<x\mid(E-H_{0}+i\epsilon)^{-1}\mid x'> & = & \frac{k}{v_{f}(4i)}[H_{0}(z)+i\ M_{\alpha}H_{1}(z)]
\label{hankel-simplify}
\end{eqnarray}
Now we return to the main idea of setting up Born scattering in the form of self consistent integral equations.
\begin{equation}
<x\mid\psi^{+}>=<x\mid\phi>+\int d^{2}x'<x\mid(E-H_{0}+i\epsilon)^{-1}\mid x'><x'\mid V\mid\psi^{+}>\label{eq:bornec1}\end{equation}
We apply the eikonal approximation to the above integral equation by assuming $\mid\vec{x}\mid\gg\mid\vec{x}'\mid$ which yields $\mid\vec{x}-\vec{x}'\mid\simeq r-\hat{r}\cdot\vec{x}'$ or $\vec{k}'=k\ \hat{r}=(k,\phi)$. In other words $\alpha\simeq\phi$, which implies that we are calculating Born scattering to the first order only. Now using this approximation, we redefine our variables as $M_{\alpha}\simeq M_{\phi}=\tau_{3}\cos(\phi)+\tau_{1}\sin(\phi)$
and $z=k\mid\vec{x}-\vec{x}'\mid=kr-\vec{k'}\cdot\vec{x'}$. Under these conditions Eq.~(\ref{hankel-simplify}) further simplifies to,
\begin{equation}
<x\mid(E-H_{0}+i\epsilon)^{-1}\mid x'>=\frac{k}{v_{f}(4i)}[H_{0}(kr-\vec{k'}\cdot\vec{x'})+i\ M_{\phi}H_{1}(kr-\vec{k'}\cdot\vec{x'})]\end{equation}
In the Born limit, $kr\gg1$, we can use the asymptotic form of the Hankel functions of first kind ($H_{n}(z)=\sqrt{\frac{2}{\pi z}}e^{i(z-\frac{\pi}{4}-n\frac{\pi}{2})}$) in the above expression to give us
\begin{equation}
<x\mid(E-H_{0}+i\epsilon)^{-1}\mid x'>=\sqrt{\frac{k\ e^{-i\frac{3\pi}{2}}}{v_{f}^{2}(8\pi)r}}[1+\ M_{\phi}]e^{ikr}\ e^{-i\vec{k'}\cdot\vec{x'}}
\label{born-asym}
\end{equation}
The incoming plane wave in the position basis can be written as $<x\mid\phi>=e^{i\,\vec{k}\cdot\vec{x}}\left(\begin{array}{c}
\cos\frac{\theta}{2}\\ \sin\frac{\theta}{2}\end{array}\right)$ where $\quad\vec{k}=(k,\theta)$ and $\mid r\mid=\mid x\mid$, as in Eq.~(\ref{eq:Phii}).
Using Eq.~(\ref{born-asym}) we have
\begin{eqnarray*}
<x\mid\psi^{+}> & = & <x\mid\phi>+\int d^{2}x'd^{2}x''<x\mid(E-H_{0}+i\epsilon)^{-1}\mid x'><x'\mid V\mid x''><x''\mid\phi>\\
 & = & e^{i\,\vec{k}\cdot\vec{x}}\left(\begin{array}{c}
\cos\frac{\theta}{2}\\
\sin\frac{\theta}{2}\end{array}\right)+\sqrt{\frac{k\ e^{-i\frac{3\pi}{2}}}{v_{f}^{2}(8\pi)r}}e^{ikr}\int d^{2}x'd^{2}x''e^{i\vec{k}\cdot\vec{x''}-i\vec{k}\cdot\vec{x'}}<x'\mid V\mid x''>[1+\ M_{\phi}]\left(\begin{array}{c}
\cos\frac{\theta}{2}\\
\sin\frac{\theta}{2}\end{array}\right)\end{eqnarray*}
The most general form of potential can be written as $V=V_{0}+V_{1}\tau_{1}+V_{2}\tau_{2}+V_{3}\tau_{3}=\sum_{n}V_{n}\tau_{n}$
where $\tau_{n}$ are the Pauli matrices and $\tau_{0}=I$. In this general representation for $V$ we have
\begin{eqnarray*}
<x\mid\psi^{+}>=\Phi(r) & = & e^{i\,\vec{k}\cdot\vec{x}}\left(\begin{array}{c}
\cos\frac{\theta}{2}\\
\sin\frac{\theta}{2}\end{array}\right)+\sqrt{\frac{k\ e^{-i\frac{3\pi}{2}}}{v_{f}^{2}(8\pi)r}}e^{ikr}\int d^{2}x'd^{2}x''e^{i\vec{k}\cdot\vec{x''}-i\vec{k}\cdot\vec{x'}}\sum_{n}<x'\mid V_{n}\mid x''>[1+\ M_{\phi}]\tau_{n}\left(\begin{array}{c}
\cos\frac{\theta}{2}\\
\sin\frac{\theta}{2}\end{array}\right)\end{eqnarray*}
We now simplify the matrices acting on the spinor
\begin{eqnarray}
[1+\ M_{\phi}]\tau_{n} & = & \left(\begin{array}{cc}
1+\cos\phi & \sin\phi\\
\sin\phi & 1-\cos\phi\end{array}\right)\tau_{n}\\
\left(\begin{array}{cc}
1+\cos\phi & \sin\phi\\
\sin\phi & 1-\cos\phi\end{array}\right)\tau_{n}\left(\begin{array}{c}
\cos\frac{\theta}{2}\\
\sin\frac{\theta}{2}\end{array}\right) & = & 2\beta_{n}(\phi,\theta)\left(\begin{array}{c}
\cos\frac{\phi}{2}\\
\sin\frac{\phi}{2}\end{array}\right)
\end{eqnarray}
where
\begin{eqnarray}
\beta_{0}=\cos(\frac{\phi-\theta}{2}), & \ \beta_{1} & =\sin(\frac{\phi+\theta}{2})\\
\beta_{2}=i\sin(\frac{\phi-\theta}{2}), & \ \beta_{3} & =\cos(\frac{\phi+\theta}{2})
\end{eqnarray}
We can write the complete wave function in a suggestive form representing plane wave and scattered wave
\begin{eqnarray}
\Phi(r) & = & e^{i\,\vec{k}\cdot\vec{x}}\left(\begin{array}{c}
\cos\frac{\theta}{2}\\
\sin\frac{\theta}{2}\end{array}\right)+\frac{e^{ikr}}{\sqrt{r}}f(\phi,\theta)\left(\begin{array}{c}
\cos\frac{\phi}{2}\\
\sin\frac{\phi}{2}\end{array}\right)
\end{eqnarray}
where
\begin{eqnarray}
f(\phi,\theta) & = & \sqrt{\frac{k\ e^{-i\frac{3\pi}{2}}}{v_{f}^{2}(2\pi)}}\int d^{2}x'd^{2}x''e^{i\vec{k}\cdot\vec{x''}-i\vec{k}\cdot\vec{x'}}\sum_{n}<x'\mid V_{n}\mid x''>\beta_{n}(\phi,\theta)
\end{eqnarray}
Here $f(\phi,\theta)$ is the scattering amplitude from which we can easily calculate the differential cross section by
\begin{equation}
\frac{d\sigma}{d\phi}=\mid\sqrt{\frac{k\ e^{-i\frac{3\pi}{2}}}{v_{f}^{2}(2\pi)}}\int d^{2}x'd^{2}x''e^{i\vec{k}\cdot\vec{x''}-i\vec{k}\cdot\vec{x'}}\sum_{n}<x'\mid V_{n}\mid x''>\beta_{n}(\phi,\theta)\mid^{2}\end{equation}
If $V$ is local and scalar we have the following properties, $<x'\mid V_{n}\mid x''>=V(x')\delta(x'-x'')$
and $V_{1}=V_{2}=V_{3}=0$.

Under these conditions the scattering amplitude simplifies to
\begin{eqnarray}
f(\varphi) & = & \sqrt{\frac{k\ e^{-i\frac{3\pi}{2}}}{v_{f}^{2}(2\pi)}}\ \cos\frac{\varphi}{2}\ \int d^{2}r\ e^{i\ \vec{q}.\vec{r}}V(\vec{r})\
 \end{eqnarray}
  where $\varphi=\phi-\theta,\ \vec{q}=\vec{k}-\vec{k'}$ and $q=2k\ \sin\frac{\varphi}{2}\ (\mid k\mid=\mid k'\mid)$ and hence
\begin{eqnarray}
\frac{d\sigma}{d\phi} & = & \mid\sqrt{\frac{k\ e^{-i\frac{3\pi}{2}}}{v_{f}^{2}(2\pi)}}\ \cos\frac{\varphi}{2}\ \int d^{2}r\ e^{i \vec{q}\cdot\vec{r}}V(\vec{r})\ \mid^{2}=\mid\sqrt{\frac{k\ e^{-i\frac{3\pi}{2}}}{v_{f}^{2}(2\pi)}}\ \cos\frac{\varphi}{2}\ \tilde{V\ }(\vec{q})\mid^{2}\label{dcs}
\end{eqnarray}
The only part that needs to be evaluated is the integral $\int d^{2}r\ e^{i\vec{q}\cdot\vec{r}}V(\vec{r})=\tilde{V\ }(\vec{q})$. The effective potential for the single vortex superflow, Eq.~(\ref{eq:ps}) is
\begin{eqnarray}
V(\vec{r}) & = & v_{f}P_{sx}=-v_{f}P_{s}(\vec{r})\sin\phi'
\end{eqnarray}
where
$P_{s}(\vec{r})=\frac{1}{2}(\frac{1}{r}-\frac{r}{R^{2}})\Theta(R-r)$ and $\sin\phi'=\sin(\varphi'+\theta')$.
It is important to note the following definitions of the angles involved in the scattering.
\begin{eqnarray*}
\vec{k} & = & (k,\theta)\ \mbox{(incoming\ momentum\ vector)}\\
\ \vec{k}' & = & (k,\phi')\ \mbox{(scattered\ momentum\ vector)}\\
\vec{q} & = & \vec{k}-\vec{k'}=(q,\theta'),\ \vec{r}=(r,\phi')\\
\varphi' & = & \phi'-\theta',\ \varphi=\phi-\theta\ \end{eqnarray*}
Now we are ready to evaluate $\int d^{2}r\ e^{i\vec{q}\cdot\vec{r}}V(\vec{r})=\tilde{V\ }(\vec{q})$
\begin{eqnarray}
\tilde{V\ }(\vec{q}) & = & -v_{f}\intop_{0}^{\infty}rdrP_{s}(\vec{r})\intop_{-\pi}^{\pi}d\varphi'\ e^{iqr\cos\varphi'}\sin(\varphi'+\theta')\\
 & = & -v_{f}\intop_{0}^{\infty}rdr\frac{1}{2}(\frac{1}{r}-\frac{r}{R^{2}})\Theta(R-r)\intop_{-\pi}^{\pi}d\varphi'\ e^{iq r\cos\varphi'}\cos\varphi'\sin\theta'\end{eqnarray}
Using some Bessel functions identities we can reduce this integral to a very compact form involving Bessel function $J_{1}$.
\begin{eqnarray}
\tilde{V\ }(\vec{q}) & = & -i\pi v_{f}\sin\theta'\frac{1}{q}(1-\frac{2}{qR}J_{1}(qR))\label{eq:ftpotbessel}\\
q\ & = & 2k\ \mid\sin\varphi/2\mid
\end{eqnarray}
Plugging Eq.~(\ref{eq:ftpotbessel}) back in Eq.~(\ref{dcs}) we get
\begin{eqnarray}
\frac{d\sigma}{d\varphi} & = & \mid\pi v_{f}\sqrt{\frac{k\ e^{-i\frac{3\pi}{2}}}{v_{f}^{2}(2\pi)}}\ \cos\frac{\varphi}{2}\ \sin\theta'\frac{1}{q}(1-\frac{1}{qR}J_{1}(qR))\mid^{2}\label{dcs1}
\end{eqnarray}
Also, note the following relations which will help simplifying Eq.~(\ref{dcs1}) further,
\begin{eqnarray}
\cos\frac{\varphi}{2}\ \sin\theta' & = & \cos\frac{\varphi}{2}(\frac{q_{y}}{q})=\cos\frac{\varphi}{2}(\frac{\sin\theta-\sin\phi}{2\mid\sin\varphi/2\mid})\\
 & = & -\frac{1}{2}(\cos(\varphi+\theta)+\cos(\theta))sgn(\varphi)
\end{eqnarray}
Therefore, simplifying Eq.~(\ref{dcs1}) further we can write the full cross section as
\begin{equation}
\frac{d\sigma}{d\varphi}=\frac{\pi\ }{32k}\frac{1}{\sin^{2}(\frac{\varphi}{2})}(1-\frac{J_{1}(2kR\ \mid\sin\varphi/2\mid)}{kR\ \mid\sin\varphi/2\mid})^{2}(\cos(\varphi+\theta)+\cos(\theta))^{2}\label{eq:finalresult}\end{equation}
The above quasiparticle scattering cross section is the contribution from one node only. We will now calculate the contributions from the other three nodes see (Eq.~(\ref{eq:4nodeaverage})). $\theta$ and $\varphi$ are the Node 1 angles and we calculate the rest of the node contribution with
respect to this node.

For Node 2 ($\theta_{2}=-(\theta-\frac{\pi}{2}),\ \varphi_{2}=-\varphi$)
\begin{equation}
\cos(\varphi_{1}+\theta_{1})+\cos(\theta_{1})=\sin(\varphi+\theta)+\sin(\theta)\label{eq:node2}\end{equation}
For Node 3 ($\theta_{3}=(\theta+\pi),\ \varphi_{3}=\varphi$)
\begin{equation}
\cos(\varphi_{3}+\theta_{3})+\cos(\theta_{1})=-\cos(\varphi+\theta)-\cos(\theta)\label{eq:node3}\end{equation}
For Node 4 ($\theta_{4}=-(\theta+\frac{\pi}{2}),\ \varphi_{4}=-\varphi$)
\begin{equation}
\cos(\varphi_{4}+\theta_{4})+\cos(\theta_{4})=-\sin(\varphi+\theta)-\sin(\theta)\label{eq:node2}\end{equation}

Now we will perform the four node average of the differential scattering cross section,
\begin{eqnarray*}
\frac{d\sigma}{d\varphi} & = & \frac{1}{4}\sum_{j=1}^{4}(\frac{d\sigma}{d\varphi})_{j}=\frac{\pi\ }{32k}\frac{1}{\sin^{2}(\frac{\varphi}{2})}(1-\frac{J_{1}(2kR\ \mid\sin\varphi/2\mid)}{kR\ \mid\sin\varphi/2\mid})^{2}<(\cos(\varphi+\theta)+\cos(\theta))^{2}>_{node-average}\end{eqnarray*}
It turns out that,
\begin{equation}
<(\cos(\varphi+\theta)+\cos(\theta))^{2}>_{node-average}=2\cos^{2}(\frac{\varphi}{2})\end{equation}
which gives us
\begin{equation}
\frac{d\sigma}{d\varphi}=\frac{\pi\ }{16k}\frac{\cos^{2}(\frac{\varphi}{2})}{\sin^{2}(\frac{\varphi}{2})}(1-\frac{J_{1}(2kR\ \mid\sin\varphi/2\mid)}{kR\ \mid\sin\varphi/2\mid})^{2}\label{eq:nodeavg}\end{equation}
We have therefore obtained a closed-form expression for $\frac{d\sigma}{d\varphi}$ from this straightforward Born-limit calculation. It agrees with our exact results in the weak potential limit.

\bibliography{apssamp}
\end{document}